\documentclass[twocolumn,aps,pre,
 amsmath,amssymb,
]{revtex4-2}

\usepackage{graphicx}% Include figure files
\usepackage{dcolumn}% Align table columns on decimal point
\usepackage{bm}% bold math
\usepackage{color}
\usepackage{hyperref}% add hypertext capabilities
\hypersetup{
    colorlinks=true,
    linkcolor=blue,
    urlcolor=blue,
    linktoc=all
}
\usepackage{float}
\usepackage{mathtools}

\DeclarePairedDelimiter\abs{\lvert}{\rvert}%

\begin{document}

\title{Optimal cost tuning of frustration: Achieving desired states in the Kuramoto-Sakaguchi model}

\author{Gemma Rosell-Tarrag\'{o} }
 \email{gemmaroselltarrago@gmail.com}
  \affiliation{Departament de F\'{i}sica de la Mat\`{e}ria Condensada, Universitat de Barcelona, Mart\'{i} Franqu\`{e}s 1, 08028 Barcelona, Spain}
\affiliation{Universitat de Barcelona Institute of Complex Systems (UBICS), Mart\'{i}  Franqu\`{e}s, 1, 08028 Barcelona, Spain
}\author{Albert D\'{i}az-Guilera}
 \email{albert.diaz@ub.edu}
  \affiliation{Departament de F\'{i}sica de la Mat\`{e}ria Condensada, Universitat de Barcelona, Mart\'{i} Franqu\`{e}s 1, 08028 Barcelona, Spain}
\affiliation{Universitat de Barcelona Institute of Complex Systems (UBICS), Mart\'{i}  Franqu\`{e}s, 1, 08028 Barcelona, Spain
}

\date{\today}

\begin{abstract}
There are numerous examples of studied real-world systems that can be described as dynamical systems characterized by individual phases and coupled in a network like structure. Within the framework of oscillatory models, much attention has been devoted to the Kuramoto model, which considers a collection of oscillators interacting through a sinus function of the phase differences. In this paper, we draw on an extension of the Kuramoto model, called the Kuramoto-Sakaguchi model, which adds a phase lag parameter to each node. We construct a general formalism that allows to compute the set of lag parameters that may lead to any phase configuration within a linear approximation. In particular, we devote special attention to the cases of full synchronization and symmetric configurations. We show that the set of natural frequencies, phase lag parameters and phases at the steady state is coupled by an equation and a continuous spectra of solutions is feasible. In order to quantify the system's strain to achieve that particular configuration, we define a cost function and compute the optimal set of parameters that minimizes it. Despite considering a linear approximation of the model, we show that the obtained tuned parameters for the case of full synchronization enhance frequency synchronization in the nonlinear model as well.
\end{abstract}

\maketitle

\section{\label{sec:introduction}Introduction}

Emergence is one of the key concepts in the analysis of complex systems \cite{nicolis2007}. Collective properties emerge as a consequence of irregular interactions among its elemental constituents \cite{barrat2008}. One of the most paradigmatic examples of emergence is synchronization \cite{Pikovsky2001,osipov2007}, because the interplay between populations of oscillatory units gives rise to a variety of global states, ranging from perfect synchronization or phase locked stationary configurations to chimera states \cite{Arenas2008,dorfler2014,boccaletti2018}. Among the different models that have been used to understand such collective behavior, a lot of effort has been devoted to the Kuramoto model (KM), in which phase oscillators interact continuously with other units through a sine function of the phase difference \cite{Kuramoto1975,Sakaguchi1986,Acebron2005}.

In the past few years there has been a growing interest in the concept of controllability, which quantifies the feasibility to achieve a desired final state of a given dynamical system \cite{liu2016}. As stated above, the KM can give rise to a wide variety of stationary (phase or frequency synchronized) or not stationary states, being chimeras an unexpected mixture of both types of behaviors \cite{frolov2020}. In this context, controllability can be understood as a tuning of the internal parameters of the oscillators to reach specific phase configurations. The most simple settings stand for a collection of identical oscillators interacting through a sinus function of the phase differences. In this case it is quite intuitive to see that the final state is a perfectly synchronized one in which all oscillators have exactly the same phase and frequency (the same frequency than the intrinsic one). It is the existence of a distribution of frequencies that gives rise to a transition, in terms of the strength of the coupling, from an incoherent state to a coherent one \cite{Kuramoto2003}; such a transition is robust in the sense that the introduction of a lag term, a phase added to the argument of the sinus function, does not change the behavior, as far as it is kept below $\pi/2$ \cite{Sakaguchi1986}. However, the introduction of this lag term for identical oscillators changes completely the structure of the, in principle, synchronized state. In Reference \cite{Nicosia2013a} it was shown that, for small and common values of the lag parameters, the synchronized state breaks into partially synchronized groups of oscillators, being symmetry the reason for the phase synchronization of the oscillators. When increasing this common lag parameter the system enters into a incoherent chaotic state. Actually, there has been an increasing interest in the last months on the role that symmetries plays in the synchronization of oscillatory units and how the lack of homogeneity in some of the parameters can be compensated by other choices \cite{Nishikawa2016,molnar2020,Zhang2020}.

In a previous work we introduced the concept of ``functionability" as a measure of the ability of a given node to change the state of the system by just tuning one internal variable, the node lag in the argument of the sinus function of the interaction \cite{Rosell-Tarrago2020}. Being an intrinsic property of the node, its change produces a global change in the phases of the system of oscillators that can be measured. Functionability stands for the reaction of the whole phase distribution to a small change in a node. The analytical expression of functionability reports its quadratic dependence on the node degree and the node lag value, but also a structural term, such that the most peripheral nodes in the network have also larger contributions to functionability centrality measure. The nodes with higher functionability values may represent positive actors for the network, because they enable more variability in the states of the system, but also potentially dangerous ones, as tiny perturbations can produce cascade like effects that completely changes the network dynamics. 

As stated, the addition of a phase lag parameter enables a richer configuration state. However, it is clear that a tuning of a single parameter will not be enough to generate the wide variety of stationary states that a population of Kuramoto oscillators can achieve. Notwithstanding, the question that arises is whether a fine tuning of a set of individual parameters can make it possible. In this paper, this is our proposal, we construct a general formalism that allows, within a linear approximation, to compute the set of lag parameters that may lead to any phase configuration for a fixed set of intrinsic frequencies. The problem can also be posed the other way around. Namely, given a set of frequencies, we may derive the configuration of phases that is produced by a set of lag parameters. 

There are numerous examples of real-world systems that can be described as dynamical systems characterized by individual phases and which functioning are object of investigation. Some examples are the brain functional networks arising from temporal correlation patterns, ac power in power grids\cite{Lei2001}, heartbeats\cite{Acharya2007}, multiprocessors and multicore processors, or traffic signaling. Not only the synchronization between their constituents may be intended or prevented, but also other particular configurations may be of relevant interest. For this reason, we propose a mechanism for tuning the intrinsic parameters of the system to achieve any desired phase configuration. 

A previous work proposes a methodology to enhance frequency synchronization for the nonlinear Kuramoto-Sakaguchi model (extension of the Kuramoto model with a node phase lag parameter) \cite{Brede2016}. Another work suggests that an unstable synchronized state becomes stable when, and only when, the oscillator parameters are tuned to nonidentical values \cite{Nishikawa2016}. We highlight the work done in Reference \cite{Skardal2014}, where the particular configuration of perfect synchronization is studied and the \textit{synchrony alignment function} is defined in order to minimize the order parameter of the system considering different topologies and frequency scenarios. We address the most general question, following a similar path to that pursued by them, forcing the system to achieve any particular configuration for the linear case of the Kuramoto-Sakaguchi model by means of a fine tuning of the phase lag or frustration parameter set. Despite considering a linear approximation of the model, we show that the obtained tuned parameters for the case of full synchronization enhance frequency synchronization in the nonlinear model as well. 

The structure of the paper is the following. In Section \ref{sec:models} we present the Kuramoto-Sakaguchi model (a variation of the Kuramoto model) as the proper framework for our purposes. Next, in Section \ref{sec:analytic}, we derive the analytic expression of the fine tuning of the frustration parameters so as to achieve any phase configuration. Then, in Section \ref{sec:energy}, we define a cost function to assess the expense of achieving a particular phase configuration by its corresponding tuning and derive the analytic solution for the cases of symmetric and fully synchronized configurations, in Sections \ref{sec:symmetric} and \ref{sec:synchronized}, respectively, as well as comment on the nonlinear validity of our results. We conclude in Section \ref{sec:conclusions}. An easy-to-follow example and further mathematical derivations can be found in the Appendix.

\section{\label{sec:KS-Model}The Kuramoto-Sakaguchi Model}
\label{sec:models}
In 1975, Kuramoto suggested one of the best-known dynamical equation to model interacting oscillatory systems\cite{Kuramoto1975}: a set of $N$ phase oscillators characterized by their phase, $\theta_i$, and coupled between each other by the sine of their phase differences. Each unit is influenced directly by the set of its nearest neighbours via the adjacency matrix of the network corresponding to the system, $\mathcal{G}(\mathcal{V}, \mathcal{E})$. The coupling strength describes the intensity of such pair-wise interactions, $K_{ij}$. The set of nodes of the network, $\mathcal{V}(\mathcal{G})$ consists of all the oscillators, while the set of edges, $\mathcal{E}(\mathcal{G})$, is made of the links between them.
In his original work, Kuramoto assumed homogeneous interactions, i.e, $K_{ij}=K \ \forall(i,j)$ \citep{Kuramoto1975,Acebron2005}. 
Taking into account the connectivity or topology of the network and the oscillatory dynamics, the dynamics of the system can be written as a system of differential equations\cite{Arenas2008}:
\begin{equation}
    \frac{d\theta_i}{dt}=\omega_i+K\sum_j A_{ij}\sin(\theta_j-\theta_i) \ i =1,...,N \ j\in \Gamma_{i}
\end{equation}
where $\Gamma_i$ is the set of neighbors of node $i$ and $\omega_i$ is the natural frequency of each unit.

We consider that two nodes are phase synchronized when their phases have the same value,
\begin{equation*}
     \theta_i(t)-\theta_j(t)=0 \ \forall t > t_0
\end{equation*}
When the phase difference has a constant value, that is, $\theta_i(t)-\theta_j(t)=c \ \forall t > t_0$, we say there is a phase locking between nodes $i$ and $j$. Similarly, we consider that two nodes are frequency synchronized when their frequencies have the same value:
\begin{equation*}
    \frac{d\theta_i}{dt}-\frac{d\theta_j}{dt}=0 \ \forall t > t_0
\end{equation*}
We say that two nodes are fully synchronized when they are phase synchronized, because this implies frequency synchronization.

As long as the distribution of natural frequencies is homogeneous, namely, all units have the same natural frequency, there is only one attractor of the dynamics: the fully synchronized state. It can be shown that, if the distribution of natural frequencies is unimodal, the system becomes frequency synchronized as long as the coupling strength is larger than a threshold value\cite{Acebron2005}.

In 1986, Kuramoto together with Sakaguchi presented a similar model which incorporated a constant phase lag between oscillators\cite{Sakaguchi1986} which can be written as follows:
\begin{equation}
    \frac{d\theta_i}{dt}=\omega_i+K\sum_j A_{ij}\sin(\theta_j-\theta_i-\alpha) \ i =1,...,N \ j\in \Gamma_{i}
\end{equation}where $\alpha$ is a homogeneous phase lag parameter.
This model has also become well-known and several variations of the model have been studied. It has been shown that, as long as $|\alpha| < \pi/2$, the system is not chaotic and a threshold value for the coupling strength exists above which the system becomes synchronized to a resulting frequency \cite{Sakaguchi1986}. In the particular case that considers homogeneous natural frequencies, i.e, $\omega_i=\omega_0 \ \forall i$, the frustration parameter, $\alpha$, forces the system to break the otherwise original fully synchronized state. However, partial synchronization is conserved for symmetric nodes in the network\citep{Nicosia2013a,Nishikawa2016}. As the frustration increases the asynchronous groups' phase move away from each other. 

We are interested here in a more general case, where the frustration parameter is not homogeneous but an intrinsic property of each unit:
\begin{equation}
\label{eq:KSmodel}
    \frac{d\theta_i}{dt}=\omega_i+K\sum_j A_{ij}\sin(\theta_j-\theta_i-\alpha_i) \ i =1,...,N \ j\in \Gamma_{i}
\end{equation}
In this context, a recent work studies a particular effect of this frustration parameter by defining functionability, a new centrality measure of the nodes in a network, in order to address the issue of which nodes, when perturbed, move the system from a synchronized state to one that is more asynchronous in the sense that it enhances the phase differences between all pairs of oscillators \cite{Rosell-Tarrago2020}. 
\section{\label{sec:analytic}Analytic expression of the frustration parameters tuning}
We address the most general problem, which considers the same dynamics as in Eq.(\ref{eq:KSmodel}), while allowing the edges of the network to be weighted, a more realistic scenario for real-world networks. 

For small values of the frustration parameters and phases close to each other, which is the case in frequency synchronization, we can linearize Eq.(\ref{eq:KSmodel}) as follows:
\begin{eqnarray}
\label{eq:KMModel}
\frac{d\theta_i}{dt}=\omega_i+K\sum_j W_{ij}(\theta_j - \theta_i -\alpha_i)=\nonumber\\=\omega_i-K\sum_j L_{ij}\theta_j -K\alpha_is_i
\end{eqnarray}
where $W_{ij}$ is the value of the weight of the edge between node $i$ and node $j$, $s_i \equiv \sum_j W_{ij}$ is the weighted degree of the $i$th node and $L$ is the weighted Laplacian matrix defined as $L_{ij} \equiv \delta_{ij} s_i - W_{ij}$.
In the stable regime, a synchronized frequency is achieved and, for all oscillators $\dot{\theta}_i = \Omega$. We can derive the value of the common frequency oscillation, $\Omega$, summing Eq.(\ref{eq:KMModel}) over $i$:
\begin{equation}
\label{eq:globalFrequencyStep1}
\sum_i \Omega = \sum_i\omega_i - K \sum_i\sum_j L_{ij}\theta_j^*-K\sum_i\alpha_is_i
\end{equation}
Taking into account the steady state $\dot{\theta}_i = \Omega \ \forall i$ and arranging summations:
\begin{equation}
\label{eq:globalFrequencyStep2}
N \Omega = \sum_i\omega_i - K \sum_j\theta_j^*\sum_i L_{ij}-K\sum_i\alpha_is_i.
\end{equation}
and finally,
\begin{equation}
\label{eq:globalFrequency}
\Omega = \left\langle \omega \right\rangle - K\left\langle \alpha s \right\rangle.
\end{equation}
where we have used the Laplacian matrix property: $\sum_i L_{ij}= 0$ and defined the averages $\sum_i\alpha_is_i/N = \left\langle \alpha s \right\rangle$ and $\sum_i\omega_i/N= \left\langle \omega \right\rangle$.
Now we can plug expression Eq.(\ref{eq:globalFrequency}) to Eq.(\ref{eq:KMModel}) to get the stable phases of oscillators, $\theta_i^*$:
\begin{equation}
\label{eq:KMStable} 
\sum_j L_{ij}\theta_j^* = \frac{\omega_i}{K}-\frac{\left\langle \omega \right\rangle}{K} + \left\langle \alpha s \right\rangle - \alpha_is_i \hspace{1cm} \forall i
\end{equation}
The solution of Eq.(\ref{eq:KMStable}) regarding phases is undetermined due to the singular nature of the Laplacian matrix. Hence, Eq.(\ref{eq:KMStable}) is, in general, an undetermined system of linear equations, that is, there is one free phase, which we should use as a reference value for the solution. 
Nonetheless, we do not work directly with the functional form of phases because they are time dependent $\{ \theta_i ^{*}\} = f_i(t)$, but with the phase differences with respect to a reference node, once the stationary state is achieved,
\begin{equation}
\label{eq:referencePhases}
\phi_i \equiv \theta_i - \theta_R
\end{equation}
In this way, we work with time independent values. In this situation, $\phi_R = 0$, by definition, as $\phi_R \equiv  \theta_R - \theta_R = 0$.

On the other hand, the contribution $ \left\langle \alpha s \right\rangle - \alpha_is_i$ of the right-hand side of Eq.(\ref{eq:KMStable}) can be written in matrix form as:
\begin{eqnarray*}
\label{eq:matrixDefinition}
-\begin{pmatrix}
\frac{N-1}{N} & -\frac{1}{N} & -\frac{1}{N} & ... \\
-\frac{1}{N} & \frac{N-1}{N} & -\frac{1}{N} & ... \\
... & ... & ... & ...\\
-\frac{1}{N} & -\frac{1}{N}1 & ... & \frac{N-1}{N} 
\end{pmatrix}\cdot\begin{pmatrix}
s_0 & 0 & ... & 0 \\
0 & s_1 & ... & 0 \\
0 & ... & ... & 0 \\
0 & ... & 0 & s_{N-1} 
\end{pmatrix}\cdot \nonumber\\ \cdot \begin{pmatrix}
\alpha_0 \\
\alpha_1 \\
... \\
\alpha_{N-1} 
\end{pmatrix}=  (-M \cdot D_s) \vec{\alpha} 
\end{eqnarray*}
where we have defined: 
$M \equiv \begin{pmatrix}
\frac{N-1}{N} & -\frac{1}{N} & -\frac{1}{N} & ... \\
-\frac{1}{N} & \frac{N-1}{N} & -\frac{1}{N} & ... \\
... & ... & ... & ...\\
-\frac{1}{N} & -\frac{1}{N}1 & ... & \frac{N-1}{N} 
\end{pmatrix}$ and $D_s \equiv \begin{pmatrix}
s_0 & 0 & ... & 0 \\
0 & s_1 & ... & 0 \\
0 & ... & ... & 0 \\
0 & ... & 0 & s_{N-1} 
\end{pmatrix}$. 
We write Eq.(\ref{eq:KMStable}) in matrix as
\begin{equation}
\label{eq:KMmatrix}
L \vec{\theta}^{*} = \frac{1}{K}\vec{\Delta \omega} - M \cdot D_s \vec{\alpha}
\end{equation}where $\Delta \omega_i \equiv \omega_i-\left\langle \omega \right\rangle$. Finally, we obtain the set of unknowns $\{\alpha_i\}$:
\begin{equation}
\label{eq:KMmatrixAlpha}
 M \cdot D_s \vec{\alpha} = \frac{1}{K}\vec{\Delta \omega}-L \vec{\theta}^{*} 
\end{equation}
Equation (\ref{eq:KMmatrixAlpha}), however, does not have a solution, because of the singular nature of $M\cdot D_s$ matrix. $M$ matrix is singular too, and hence, its inverse matrix does not exist. Mathematically, $det(M\cdot D_s)=det(M)\cdot det(D_s)=0$

Similarly as we did for phases \citep{Rosell-Tarrago2020}, we solve the singularity problem by setting a reference node, which we call \textit{control} node, regarding frustration parameters, i.e., we would not obtain the value for each of the parameters, but a relation between them:
\begin{equation}
\label{eq:referenceAlpha}
\kappa_i \equiv \alpha_i - \alpha_C
\end{equation}
where $\alpha_C$ is the value of the control node. In this situation, $\kappa_C = 0$, by definition, as $\kappa_C \equiv  \alpha_C - \alpha_C = 0$. 

To easily write the matrix expressions, we define the selection matrix $J_{(n,m)}$, which is, in general, an $(N-1) \times (N-1)$ identity matrix after the removal of the $mth$ row and the $nth$ column.

$L\vec{\theta}^{*}$ turns to $\tilde{L}(k,R) \vec{\phi}^{*}$, where we have removed the $kth$ row and the $Rth$ column. The result does not depend on which row we remove, hence we can choose any $k$. Using the selection matrix, $\tilde{L}(k,R) = J_{(,k)} \cdot L \cdot J_{(R,)} \equiv \tilde{L}$. 

Similarly, $\vec{\tilde{\phi}}(k) =J_{(,k)} \cdot \vec{\phi} \equiv \vec{\tilde{\phi}}$, where we have removed the $kth$ row.

In an equivalent way as the definition of the reduced Laplacian:
\begin{equation*}
\label{eq:reducedMDs}
\tilde{MD_s}(k, C) = J_{(,k)} \cdot MD_s \cdot J_{(C,)}  \equiv \tilde{MD_s} 
\end{equation*}
where $\tilde{MD_s}$ is $MD_s$ without the $kth$ row and the $Cth$ column.

Similarly, $\vec{\tilde{\kappa}}(k) = J_{(,k)} \cdot \vec{\kappa}\equiv \vec{\tilde{\kappa}}$ and $\vec{\tilde{\Delta \omega}}(k) =  J_{(,k)} \cdot \vec{\Delta\omega}\equiv \vec{\tilde{\Delta \omega}}$, where we have removed the $kth$ row.

Considering all the previous definitions and remarks, Eq.(\ref{eq:KMmatrix}) can be rewritten as:
\begin{eqnarray}
\label{eq:kappaSolutionComplete}
\tilde{ MD_s} \vec{\tilde{\kappa}} = \frac{1}{K}\vec{\tilde{\Delta \omega}}-\tilde{L} \vec{\tilde{\phi}}^{*}-\alpha_C \cdot J_{(,k)} \sum_j^{\rightarrow}[MD_s]_{ij} 
\end{eqnarray}
and finally,
\begin{eqnarray}
\label{eq:kappaSolutionCompleteFinal}
 \vec{\tilde{\kappa}} = \left( \tilde{ MD_s}\right)^{-1} \left(\frac{1}{K}\vec{\tilde{\Delta \omega}}-\tilde{L} \vec{\tilde{\phi}}^{*}-\alpha_C \cdot 	\tilde{M}\vec{s}\right)
\end{eqnarray}
where we have used $J_{(,k)} \vec{\sum}_j[MD_s]_{ij}=\tilde{M}\vec{s}$.
Notice that $MD_s$ matrix is singular, but the row sum is not zero, although it is so for the column sum. Hence, we need to set $\alpha_C = 0$ if we want to avoid extra constant arrays in the final expression. In this particular case:
\begin{eqnarray}
\label{eq:kappaSolution}
 \vec{\tilde{\kappa}} = \left( \tilde{ MD_s}\right)^{-1}  \left(\frac{1}{K}\vec{\tilde{\Delta \omega}}-\tilde{L} \vec{\phi}^{*} \right)
\end{eqnarray}
and keep in mind that $\kappa_C=0$.

The obtained values of $\vec{\alpha}$ depend on both the chosen control node, $C$, and the value we set for its frustration parameter, $\alpha_C$. Notice, therefore, that there is a continuous spectrum of values for the frustration parameter in order to achieve a particular phase configuration.

Moreover and more importantly, due to the non-row-sum equal to zero of $MD_s$ matrix, the differences between the obtained values are dependent of the control node choice. Mathematically, $\alpha_i-\alpha_j(C=l) \neq \alpha_i-\alpha_j(C=k)$ if $l \neq k$. This property will lead us to the definition of a cost for the system to move to the final configuration, which will depend on both the control node and the value of its frustration parameter.

\begin{figure}[H]
\centering
\includegraphics[width = 0.45\textwidth]{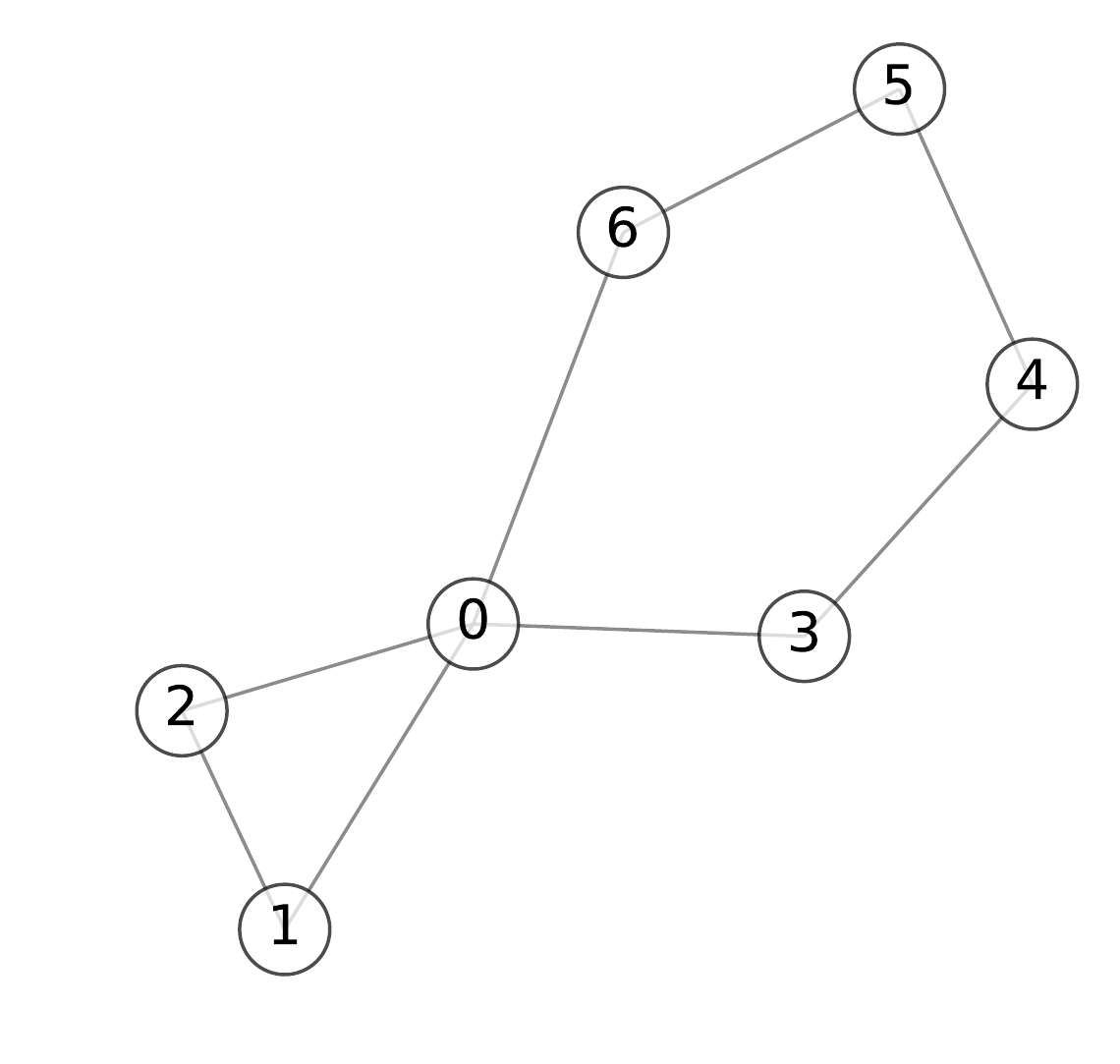}
\caption{Network of seven nodes.}
\label{fig:sevenNetwork}
\end{figure}

We provide an example of a toy network for the case of a homogeneous natural frequencies distribution, i.e., $\omega_i = \omega \ \forall i$. In this case, Eq.(\ref{eq:kappaSolutionCompleteFinal}) turns to:
\begin{equation*}
\label{eq:kappaSolutionCompleteHomogenousFrequency}
\vec{\tilde{\kappa}} =\left( \tilde{ MD_s}\right)^{-1} \left(-\tilde{L} \vec{\tilde{\phi}}^{*}-\alpha_C \cdot 	\tilde{M}\vec{s}\right)
\end{equation*} which in the case of the network depicted in Figure \ref{fig:sevenNetwork}, 
leads to the solution
\begin{equation}
\begin{pmatrix}
\kappa_0 \\
\kappa_2 \\
\kappa_3 \\
\kappa_4 \\
\kappa_5 \\
\kappa_6
\end{pmatrix} =  \begin{pmatrix}
\frac{-2	\alpha_1 + 3\phi_1+\phi_3+3\phi_6}{4}\\
\frac{3(\phi_1-\phi_2)}{2}\\
\frac{2\phi_1-\phi_2-2\phi_3+\phi_4}{2}\\
\frac{2\phi_1-\phi_2+\phi_3-2\phi_4+\phi_5}{2}\\
\frac{2\phi_1-\phi_2+\phi_4-2\phi_5-\phi_6}{2}\\
\frac{2\phi_1\phi_2+\phi_5-2\phi_6}{2}
\end{pmatrix}
\end{equation}
where we have chosen $\kappa_1 = 0$ and $\phi_0 = 0$. Hence, the results are written as a function of the value $\alpha_1$ and $\phi_i \ i \neq 0$. Therefore, we can achieve any phase configuration, given by the set $\{\phi_i\}$ by tuning the frustration parameters set $\{\alpha\}$, where $\alpha_i = \kappa_i + \alpha_C$.

To illustrate how we obtain the final values, let us consider the following phase configuration:
\begin{equation}
\vec{\tilde{\phi}}_{(R=0)} = (0.1,0.2,0.25,-0.2,-0.1,0.0)
\label{eq:example-phases}
\end{equation}
In the general case where $\alpha_C = \alpha_1 \neq 0$:
\begin{equation*}
   \vec{\tilde{\kappa}}_{(C=1)} =(0.1375-\frac{\alpha_1}{2},-0.15,-0.35,0.275,0.0,-0.05).
\end{equation*}
If we choose $\alpha_C = 0$, then $\alpha_i = \kappa_i$, we can include the value of the control node $C = 1$:
\begin{equation*}
\vec{\tilde{\alpha}} =(0.1375,0.0,-0.15,-0.35,0.275,0.0,-0.05).
\end{equation*}
Alternatively, we can choose whatever value we need regarding the control node. For instance, if $\alpha_C = \alpha_1 = 0.1$:
\begin{equation*}
\vec{\tilde{\alpha}} =(0.1875,0.1,-0.05,-0.25,0.375,0.1,0.05)
\end{equation*}
 and the phases configuration is the same.
Importantly, we recover the same phase differences using the nonlinear model with the tuned $\alpha$'s, up to an error. For this last example and using the frustration parameters obtained by setting $\alpha_1= 0.1$, the nonlinear model leads to final phases vector 
\begin{equation}
\begin{split}
\vec{\tilde{\phi}}_{(R=0)} = (0.09969,0.19944,0.25097,\\
-0.19798,-0.09897,0.00012)
\end{split}
\end{equation}which represents $\sim 0.3 \%$ of relative error with respect to the initial Eq.(\ref{eq:example-phases}). See the full derivation of the analytical solution in Appendix \ref{ap:example}.

\section{\label{sec:energy}Optimal Cost tuning of frustration}
As pointed out in Section \ref{sec:analytic}, there is a continuous spectrum of values for the choice of the frustration parameters that enables the system access a particular phase configuration. The following question arises naturally: Among all the possible solutions, which is the one that makes the system achieve a particular phase configuration with the minimum required cost?

This question is of particular relevance when we consider the plausible real nature of the system. If a real system needs to access a particular phase configuration, which may be associated with a precise function, it will tend to minimize the effort or cost to do so.

In order to quantify the required cost, we define it as follows:
\begin{equation}
\label{eq:energy}
e_T(C) \equiv \sum_i |\alpha_i(C)|
\end{equation}
Henceforth, the cost associated to each node is given by the absolute value of the required frustration parameter. The absolute value operator allows for a sign-free contribution of each node, a very convenient choice in the case that the system is not beforehand specified, and a general definition is proposed instead. Furthermore, unlike other nonlinear cost functions such as the square sum of the parameters, no extra weight is given to larger values, besides the corresponding to a linear function. 

As previously remarked, $e_T(C)$ will depend both on the chosen control node, $C$, as well as the particular choice of its frustration parameter, $\alpha_C$.

The optimal configuration is given by the solution of the minimization problem

\begin{equation}
\label{eq:minimizeEnergy_s_C}
\min_{C,x} e_T(C,x) = \min_{C,x} \sum_i^{N}  \alpha_i(C,x)
\end{equation}
where the $x$ variable is not yet specified. Depending on the problem we are interested in we would set it either to $\omega_i$, $s_i$ or any other combination of the parameters of the model. The minimal value of the cost will depend on the proper choice of the control node, $C$, in addition of the particular value of its frustration parameter, $\alpha_C$, as the free parameter left to be set. In Sections \ref{sec:symmetric} and \ref{sec:synchronized} we provide a thorough analysis of it.

\begin{figure}[H]
\centering
\includegraphics[width = 0.49\textwidth]{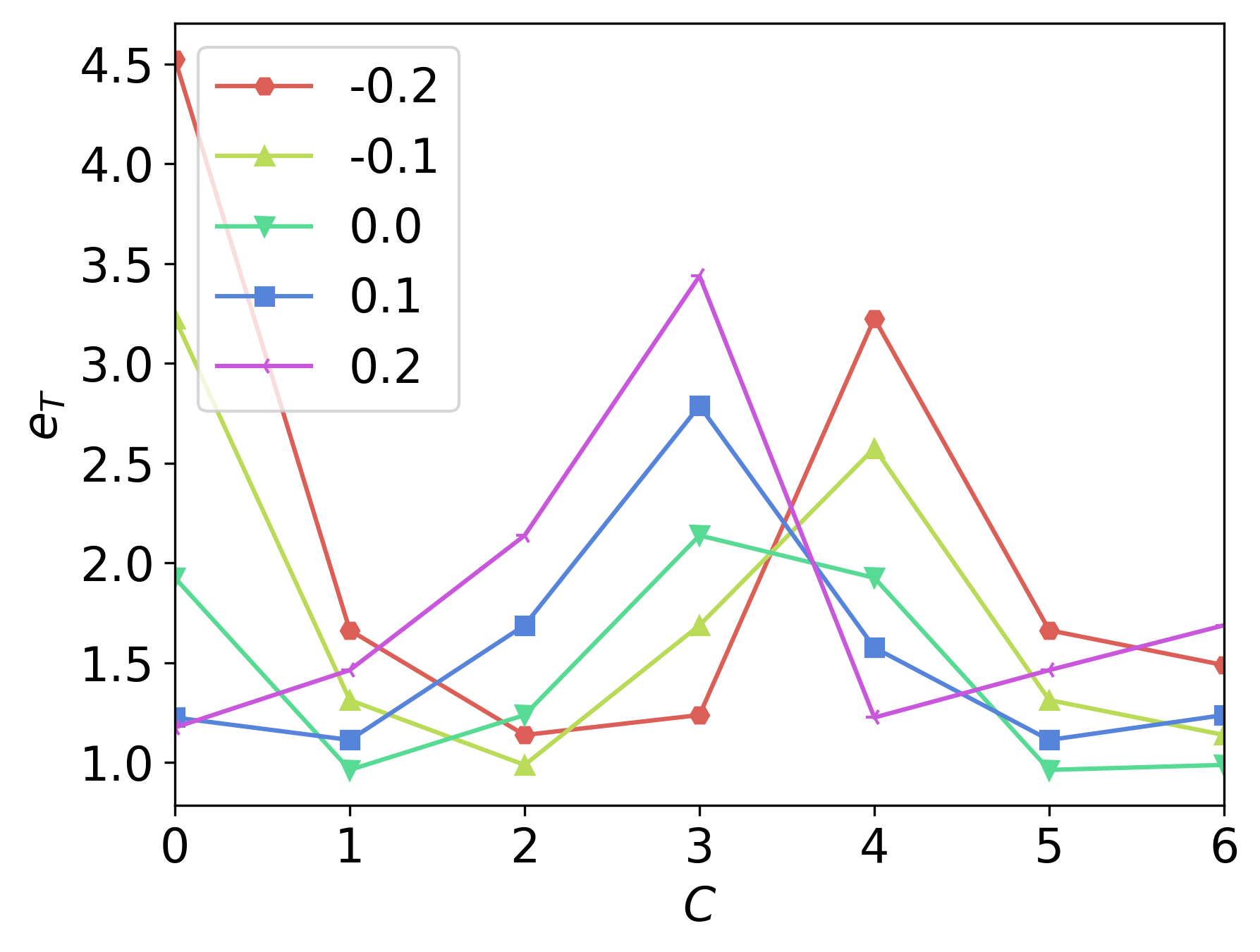}
\caption{(Color online) Implied cost to achieve the phase configuration in Eq.(\ref{eq:example-phases}) as a function of the chosen control node, $C$, for the network in Figure \ref{fig:sevenNetwork} and considering five different values of $\alpha_C$. Notice that the minimum cost is given, in this case, by $\alpha_C=0$ and $C=1$ or $C=5$.}
\label{fig:sevenEnergy}
\end{figure}

The cost required to achieve a particular phase configuration depends on that configuration, the control node and the chosen value of $\alpha_C$.
In Figure \ref{fig:sevenEnergy} we present an example, following with the network presented in Section \ref{sec:analytic} and choosing different values of $\alpha_C$, we compute numerically the values of the required cost using Eq.(\ref{eq:energy}) to achieve the phase configuration given in Eq.(\ref{eq:example-phases}). Notice that the global minimum depends on the control node and its frustration parameter.
In Section \ref{sec:analytic} we have derived the general analytical solution of the frustration parameters as a function of a particular choice for the phase configuration. In this section we have defined a cost function in order to assess the optimal choice of such configuration.

Depending on the phase configuration one is interested in achieving, results will vary and the analytical expressions will have different features.

In the following sections we will focus on two particular configurations, due to its intrinsic importance, in order to obtain and discuss the analytical solution of Eq.(\ref{eq:minimizeEnergy_s_C}): The configuration given by the symmetries of the network\cite{Nicosia2013a} and the fully synchronized state.
\section{\label{sec:symmetric}Symmetric phase configuration}
As explained in Section \ref{sec:models}, a particular example of the Kuramoto-Sakaguchi model is the symmetric case, obtained by a homogeneous distribution of frustration parameters, i.e, $\alpha_i = \alpha_h \ \forall i$. For our purposes, we consider $\alpha_h>0$. In this situation, the trivial solution of the frustration parameters, $\alpha_i = \alpha_h$, is another one of the values out of the continuous spectrum. That is, we can recover the landscape given by the symmetric configuration in many different ways. We are however, interested in computing the analytical expression of the cost function in order to select the one corresponding to the minimum cost.
\subsection{Optimal cost tuning when $\alpha_C = 0$}
\label{subsec:symm_alpha0}
Let us firstly consider the case where $\alpha_C=0$ and homogeneous natural frequencies $\omega_i = \omega_h \ \forall i$. In the particular case of the symmetric configuration, that is, the phase configuration given by $\alpha_i = \alpha_h \ \forall i$ the solution of the frustration parameters is given by:
\begin{equation}
\label{eq:kappaSolutionOmegaConstant}
\vec{\tilde{\kappa}} = \left( \tilde{ MD_s}\right)^{-1}  \left(\frac{1}{K}\vec{\tilde{\Delta \omega}}-\tilde{L} \vec{\tilde{\phi}}^{*} \right) = - \left( \tilde{ MD_s}\right)^{-1}\tilde{L} \vec{\tilde{\phi}}^{*}
\end{equation}
But $\vec{\tilde{\phi}}^{*}$ corresponds to the symmetric case. Hence (see Section \ref{sec:models}),
\begin{equation}
\label{eq:symmetricPhases}
\vec{\tilde{\phi}}^{*} = \alpha_h \tilde{L}^{-1} \vec{\tilde{\Delta s}}
\end{equation}
where $\vec{\tilde{\Delta s}}_i \equiv \left\langle s \right\rangle -s_i$ and the tilde touches on $kth$ row removal.

Plugging Eq.(\ref{eq:symmetricPhases}) into Eq.(\ref{eq:kappaSolutionOmegaConstant}):
\begin{equation*}
\vec{\tilde{\kappa}} =-\alpha_h \left( \tilde{ MD_s}\right)^{-1}\tilde{L} \tilde{L}^{-1} \vec{\tilde{\Delta s}} = -\alpha_h \left( \tilde{ MD_s}\right)^{-1} \vec{\tilde{\Delta s}}
\end{equation*}
But $\vec{\tilde{\Delta s}}$ can be written as:
\begin{equation}
\vec{\tilde{\Delta s}} = - \tilde{M} \vec{s}
\end{equation}
Putting it all together:
\begin{equation}
\label{eq:finalKappasSymmetricStep1}
\vec{\tilde{\kappa}} =-\alpha_h \left( \tilde{ MD_s}\right)^{-1}\tilde{L} \tilde{L}^{-1} \vec{\tilde{\Delta s}} = \alpha_h \left( \tilde{ MD_s}\right)^{-1} \tilde{M} \vec{s} \\ \
\vec{\tilde{\kappa}}
\end{equation}
which in vector form is written as:
\begin{equation}
\label{eq:finalKappasSymmetric}
\vec{\tilde{\kappa}} = \alpha_h \left( \tilde{ MD_s}\right)^{-1} \tilde{M} \vec{s} \\ \
\vec{\tilde{\kappa}} = \alpha_h \begin{pmatrix}
1-\frac{s_C}{s_0} \\
1-\frac{s_C}{s_1} \\
\cdots \\
1-\frac{s_C}{s_{N-1}} \\
\end{pmatrix} 
\end{equation}
And considering the relation between $\alpha$ and $\kappa$, in Eq.(\ref{eq:referenceAlpha}):
\begin{equation}
\label{eq:finalAlphasSymmetric}
\begin{split}
\vec{\alpha} = \alpha_h \begin{pmatrix}
1-\frac{s_C}{s_0} \\
1-\frac{s_C}{s_1} \\
\cdots \\ 
0 \text{ ($C$ node)}\\
\cdots\\
1-\frac{s_C}{s_{N-1}} \\
\end{pmatrix}
\end{split}
\end{equation}
Equation (\ref{eq:finalKappasSymmetric}) gives us the tuned values of the frustration parameters as a function of the chosen control node, $C$, when $\alpha_C=0$. Notice that the result depends nonlinearly only on the ratio between the degree of each node and the control node. This informs us that nodes with the same degree would be tuned to the same value or, in other words, the tuning depends only on the degree sequence of the network.

Once we have computed the analytical solution of the frustration parameters, we derive the expression of the required cost to achieve such state with the particular choice of $C$. Using the definition in Eq.(\ref{eq:energy}):
\begin{equation}
\label{eq:energySymmetric}
 e_T(C)= \abs*{\alpha_h} \sum_i^{N-1} \abs*{1-\frac{s_
 C}{s_i}} = |\alpha_h| \sum_i^{N-1} \abs[\Big]{\frac{s_
 i -s_C}{s_i}}
\end{equation}

Before we provide the mathematical solution to the minimization problem defined in Eq.(\ref{eq:minimizeEnergy_s_C}) for this particular case, let us gain an intuitive understanding of it. Looking at Eq.(\ref{eq:energySymmetric}) we see that the contribution of the $ith$ node to the cost increment depends on $|s_C-s_i|$ and, hence, if the chosen control node, $C$, has an extreme value, i.e, $s_C \ll s_i$ or $s_C\gg s_i$, the contribution will be larger. On the contrary, if the degree of the control node is similar to that of the remaining nodes, then the increase in cost will be smaller.

\begin{figure}[H]
    \centering
    \includegraphics[trim=0.35cm 0 0 0, clip,width=0.45\textwidth]{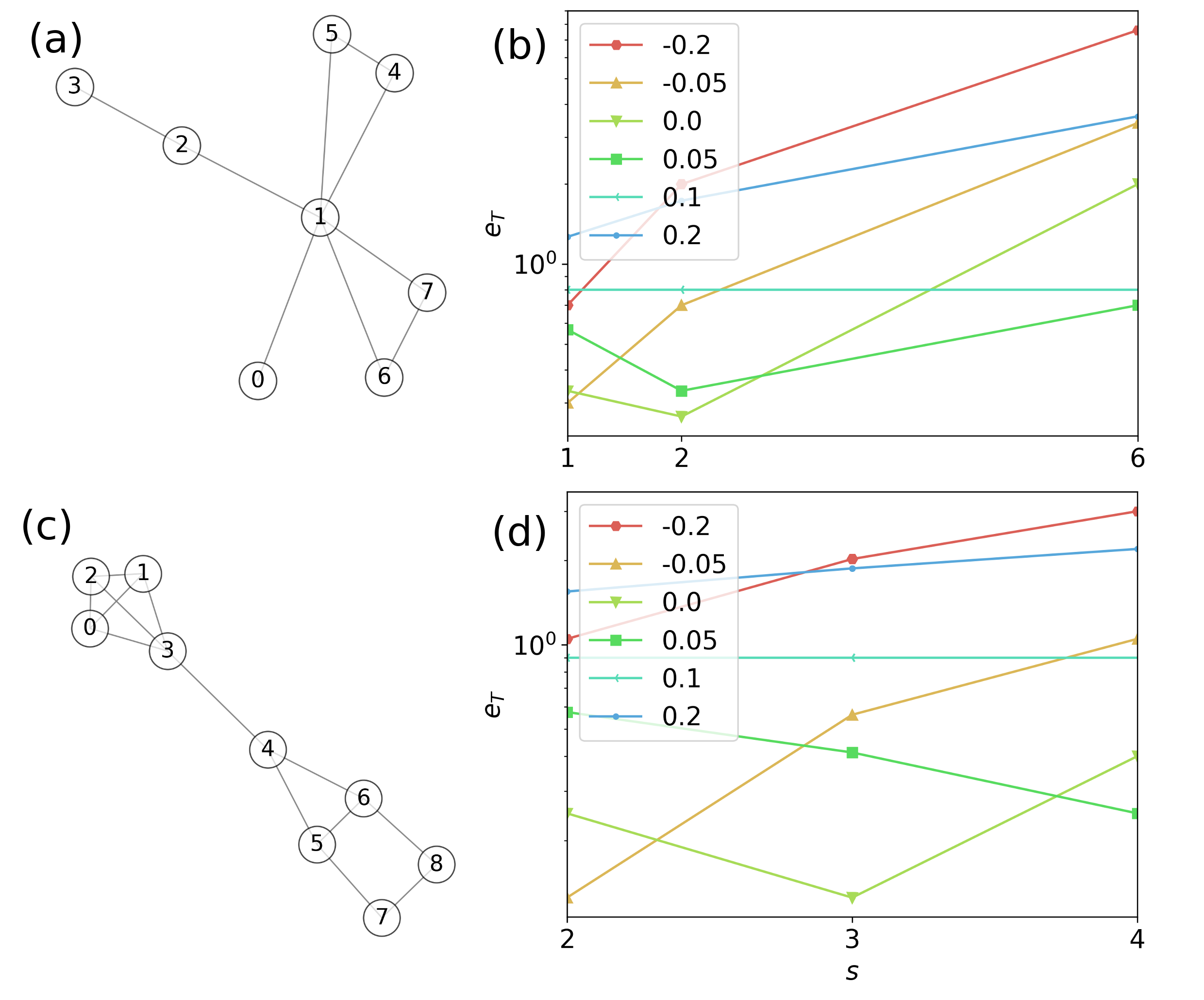}
    \caption{(Color online) Implied cost to achieve the symmetric configuration as a function of the degree corresponding to different choices of the control node, $s_C$, for a network of 8 nodes (upper panels) and 9 nodes (lower panels). The distinct colors and markers correspond to different values of $\alpha_C$. The symmetric configuration is generated by a value of $\alpha_h = 0.1$.}
    \label{fig:symmetric-energy}
\end{figure}

For example, the network in Figure \ref{fig:symmetric-energy}(a), with $\vec{s}=(1,6,2,1,2,2,2,2)$ has the set of unique degrees $\vec{s}_{unique}=(1,2,6)$ and hence three possible values of the cost, shared by some nodes. 
If $C=\{0,3\}$, $s_C=1$:
\begin{eqnarray*}
 e_T(C) = |\alpha_h| \left( |1-\frac{1}{1}| + 5|1-\frac{1}{2}|+|1-\frac{1}{6}| \right) =\nonumber\\= |\alpha_h| \left( \frac{5}{2}+\frac{5}{6} \right)=\frac{10}{3}|\alpha_h|
\end{eqnarray*}
If $C=\{2,4,5,6,7\}$, $s_C=2$:
\begin{eqnarray*}
 e_T(C) = |\alpha_h| \left( 2|1-\frac{2}{1}| + 4|1-\frac{2}{2}|+|1-\frac{2}{6}| \right) =\nonumber\\= |\alpha_h| \left(2+\frac{2}{3} \right)=\frac{8}{3}|\alpha_h|
\end{eqnarray*}
And, finally, if $C=1$, $s_C=6$:
\begin{eqnarray*}
 e_T(C) = |\alpha_h| \left( 2|1-\frac{6}{1}| + 5|1-\frac{6}{2}| \right) =\nonumber\\= |\alpha_h| \left(10+10 \right)=20 |\alpha_h|
\end{eqnarray*}
The minimum value of the energy is $\frac{8}{3}|\alpha_h|$, corresponding to the choice $C\in\{2,4,5,6,7\}$ with $s_C=2$.

Notice that the optimal choice of the control node (or nodes) does not depend on the value of $\alpha_h$ in the symmetric configuration, but only on the degree sequence of the network. Moreover, this example illustrates that the degree of the control node corresponds to an intermediate value within the degree sequence of the network and not an extreme value. A more detailed inspection of Eq.(\ref{eq:energySymmetric}) discloses that the proper choice of the control node (or control nodes) corresponds to the minimization of the relative error of degrees. In order to find the particular value of the degree that the control node must have we should solve the minimization problem defined in Eq.(\ref{eq:minimizeEnergy_s_C}):
\begin{equation*}
\min_{C,x} e_T(C,x) = \min_{C,x} \sum_i^{N}  \alpha_i(C,x)
\end{equation*}
which, when considering the symmetric configuration case, turns to
\begin{equation}
    \min_{s_C}|\alpha_h|\sum_i^{N-1}\abs[\Big]{1-\frac{s_C}{s_i}} =  |\alpha_h|\min_{s_C}\sum_i^{N}\abs[\Big]{\frac{s_C-s_i}{s_i}}
    \label{eq:minimizationSymmetric}
\end{equation}
Equation (\ref{eq:minimizationSymmetric}) is equivalent to the minimization of the absolute value of the relative error of the degree:
\begin{equation}
    |\alpha_h|\min_{s_C}\sum_i^{N}|\mathcal{E}_i|
    \label{eq:minimizationRelativeError}
\end{equation}where $\mathcal{E}_i=\abs[\Big]{\displaystyle\frac{s_C-s_i}{s_i}}$. 

The most general minimization problem of the relative error of a variable \cite{Semsar-Kazerooni2009} can be written as
\begin{equation}
    \min_d \sum_{i=1}^N w_i |x_i-d| \ ; d > 0
    \label{eq:WARE}
\end{equation}where $d$ is the variable one is interested in and $w_i$ is the weight corresponding to each $x_i$ variable. The solution of Eq.(\ref{eq:WARE}) is given by
\begin{eqnarray}
\label{eq:WARESolution}
    d = x_{m} \text{ , where } m \equiv \min \{ i \ | \ \sum_{k=1}^i w_k \geq \sum_{k=i}^n w_k \} \nonumber\\
    i \in \{1,...,n \}
\end{eqnarray}
In other words, the value of $d$ that minimizes Eq.(\ref{eq:WARE}) corresponds to the weighted median of the variable $x$ or, equivalently, the 50\% weighted percentile. The weighted median of a set $n$ distinct ordered elements $x_1,x_2,...,x_n$ with positive weights $w_1,w_2,...,w_n$, is the element $x_k$ satisfying $\min \{ i | \sum_{k=1}^i w_k \geq \sum_{k=i}^n w_k \}$. In other words, the solution is given by $x_k$, the value such that the sum of the weights at each side of the pivot, $k$, are as even as possible.

The particular case defined in Eq.(\ref{eq:minimizationSymmetric}) can be mapped to the most general problem defined in Eq.(\ref{eq:WARE}), choosing $w_i=1/s_i$, $x_i=s_i$ and $d=s_C$. Accordingly, the solution of $s_C$ corresponds to the weighted median of the set $\{ s_i \}$, with weights given by the inverse of the node degree.

Following the example of the network in Figure \ref{fig:symmetric-energy}(a), with degree sequence $\vec{s}=(1,6,2,1,2,2,2,2)$, let us compute the optimal value of $s_C$ by using Eq.(\ref{eq:WARESolution}).
\begin{eqnarray*}
\text{sorted}(\vec{s})=(1,1,2,2,2,2,2,6) \nonumber \\  \vec{w}=\left(1,1,\frac{1}{2},\frac{1}{2},\frac{1}{2},\frac{1}{2},\frac{1}{2},\frac{1}{6}\right)
\end{eqnarray*}
To find the weighted median, we have to find the minimum value such that the sum of the weights at each side of the pivot are as even as possible.
\begin{equation*}
    1+1+\frac{1}{2}+\frac{1}{2} = 3 \geq 2.17 = \frac{1}{2} + \frac{1}{2}+ \frac{1}{2}+ \frac{1}{2}+\frac{1}{6}
\end{equation*}
which corresponds to $s_C=2$, in agreement with the location of the minimum for $\alpha_C=0$ in Figure \ref{fig:symmetric-energy}(b) corresponding to the network in Figure \ref{fig:symmetric-energy}(a).
\subsection{Optimal cost tuning when $\alpha_C \neq 0$}
\label{subsec:symm_alphaNon0}
We next ask which is the optimal choice of the control node in the case we let $\alpha_C \neq 0$ and $\omega_i = \omega_h \ \forall i$. In this case, we should look at Eq.(\ref{eq:kappaSolutionComplete}) and set $\vec{\tilde{\Delta \omega}} = 0$. 
Making use of the analytical solution of the symmetric configuration in Eq.(\ref{eq:symmetricPhases}):
\begin{equation*}
\tilde{ MD_s} \vec{\tilde{\kappa}} = -\alpha_h \tilde{L}\tilde{L}^{-1} \vec{\tilde{\Delta s}} -\alpha_C \cdot \sum_j^{\rightarrow}[MD_s]_{ij}
\end{equation*}
Using the properties $\tilde{L}\tilde{L}^{-1} = \mathbb{I}$ and $\vec{\tilde{\Delta s}}=\tilde{M} \vec{s}$,
\begin{equation*}
\vec{\tilde{\kappa}} = \alpha_h \left( \tilde{ MD_s}\right)^{-1} \left( \tilde{M} \vec{s} - \alpha_C  \left( \tilde{ MD_s}\right)^{-1} \sum_j^{\rightarrow}[MD_s]_{ij} \right)
\end{equation*}
Finally, in vector form, \begin{eqnarray}
\label{eq:finalKappasCompleteStep1}
\vec{\tilde{\kappa}} =  \alpha_h \begin{pmatrix}
1-\frac{s_C}{s_0} \\
1-\frac{s_C}{s_1} \\
\cdots \\
1-\frac{s_C}{s_{N-1}} \\
\end{pmatrix} - \alpha_C \begin{pmatrix}
1-\frac{s_C}{s_0} \\
1-\frac{s_C}{s_1} \\
\cdots \\
1-\frac{s_C}{s_{N-1}} \\
\end{pmatrix}
\end{eqnarray}
Using Eq.(\ref{eq:referenceAlpha}),
\begin{eqnarray}
\label{eq:finalKappasComplete}
\vec{\alpha} = (\alpha_h-\alpha_C) \begin{pmatrix}
1-\frac{s_C}{s_0} \\
1-\frac{s_C}{s_1} \\
\cdots \\ 
0 \text{ ($C$ node)}\\
\cdots\\
1-\frac{s_C}{s_{N-1}} \\
\end{pmatrix} + \alpha_C
\end{eqnarray}
where we have used the result in Eq.(\ref{eq:finalKappasSymmetric}) and the relation
\begin{eqnarray*}
\left( \tilde{ MD_s}\right)^{-1} \sum_j^{\rightarrow}[MD_s]_{ij} = \left( \tilde{ MD_s}\right)^{-1} \tilde{M} \vec{s} = \nonumber \\= \begin{pmatrix}
1-\frac{\alpha_C}{\alpha_0} \\
1-\frac{\alpha_C}{\alpha_1} \\
\cdots \\
1-\frac{\alpha_C}{\alpha_{N-1}} \\
\end{pmatrix}
\end{eqnarray*}
In the particular case that $\alpha_C = \alpha_h$ we recover the trivial initial configuration $\alpha_i = \alpha_h \ \forall i$, as expected from the model.

Once we have computed the analytical solution of the frustration parameters, we derive the expression of the implied cost to achieve such state with the particular choice of $C$. Using the definition in Eq.(\ref{eq:energy}):
\begin{equation}
\label{eq:energySymmetricComplete}
 e_T(C) = \sum_{i=0}^{N-1} \abs[\Big]{(\alpha_h-\alpha_C)\left(1-\frac{s_C}{s_i}\right)+\alpha_C}
\end{equation}
We next derive the analytical solution of the optimal choice of the control node and finally proof that the global minimum corresponds to a value of $\alpha_C=0$.
%%%%%%%%%%%%%%%%%%%%%%%%%%%%
Equation (\ref{eq:energySymmetricComplete}) can be rearranged as
\begin{eqnarray}
 e_T(C) =  \sum_{i=0}^{N-1} \abs[\Big]{\frac{s_i-\left(1-\frac{\alpha_C}{\alpha_h}\right)s_C}{s_i/\alpha_h}}
 \label{eq:energySymmetricRearranged}
\end{eqnarray}
and thereby can be easily mapped to the solution of the minimization problem defined and solved in Eq.(\ref{eq:WARE}) and Eq.(\ref{eq:WARESolution}), respectively. Looking at Eq.(\ref{eq:energySymmetricComplete}), we should choose $x_i=s_i$, $w_i=\alpha_h/s_i$ and $d=\left(1-\alpha_C/\alpha_h\right)s_C$. With this choice, the value of $d$ that minimizes Eq.(\ref{eq:energySymmetricComplete}) corresponds to the weighted median of the set $\{s_i\}$ with weights $\alpha_h/s_i$. Therefore, the value of $d$ is the same as the solution of the case $\alpha_C = 0$, but $d \neq s_C$ and thus we must apply a transformation in order to obtain the optimal choice of $s_C$.  We have to distinguish several cases, considering $\alpha_h>0$:

\begin{itemize}
    \item $\alpha_C>0$. In this case we inspect Eq.(\ref{eq:energySymmetricRearranged}) and distinguish two more cases:
        \begin{itemize}
            \item[o] $\alpha_C>\alpha_h$: in this case, the prefactor of $s_C$ is negative, and we can write: 
            \begin{eqnarray*}
                  e_T(C) =  \sum_{i=0}^{N-1} \abs[\Big]{\frac{s_i+\abs[\Big]{1-\frac{\alpha_C}{\alpha_h}}s_C}{s_i/\alpha_h}} = \nonumber \\ = \sum_{i=0}^{N-1}\abs[\Big]{\alpha_h+M \alpha_h\frac{s_C}{s_i}}
            \end{eqnarray*}where $M \equiv \abs[\Big]{1-\frac{\alpha_C}{\alpha_h}} >0$ is a positive number. Hence, as the cost function increases with increasing $s_C$, the minimum is achieved when $s_C=\min (s_i)$ (See Figure \ref{fig:symmetric-energy} at $\alpha_C=0.2$).
            
            \item[o]$\alpha_C<\alpha_h$: in this case, the prefactor of $s_C$ is positive, and we can write:
            \begin{equation*}
                  e_T(C) =  \sum_{i=0}^{N-1} \abs[\Big]{\frac{s_i-\abs[\Big]{1-\frac{\alpha_C}{\alpha_h}}s_C}{s_i/\alpha_h}}
            \end{equation*}
            taking into account that $d=\abs[\Big]{1-\frac{\alpha_C}{\alpha_h}}s_C$ and considering that, in this case, $0< \alpha_C < \alpha_h$ and hence $0 \leq \abs[\Big]{1-\frac{\alpha_C}{\alpha_h}} \leq 1$ and the weighted mean is bounded by $\min(s_i)\leq d \leq \max(s_i)$, the optimal value of $s_C$ falls in the range $d \leq s_C \leq \max(s_i)$. Hence, the optimal value of $s_C$ is always larger than the weighted median, $d$ (see Figure \ref{fig:symmetric-energy} at $\alpha_C=0.05$).
        \end{itemize}
    \item $\alpha_C<0$. In this case we can rewrite Eq.(\ref{eq:energySymmetricRearranged}) as
    \begin{eqnarray*}
     e_T(C) =  \sum_{i=0}^{N-1} \abs[\Big]{\frac{s_i-\left(1+\frac{|\alpha_C|}{\alpha_h}\right)s_C}{s_i/\alpha_h}}
    \end{eqnarray*} and distinguish two more cases:
        \begin{itemize}
            \item[o] $|\alpha_C| > |\alpha_h|$: In this case, the prefactor of $s_C$ is positive and bounded by $2 \leq \left(1+\frac{|\alpha_C|}{\alpha_h}\right) \leq \infty$.
            In this case, $d = \left(1+\frac{|\alpha_C|}{\alpha_h}\right)s_C$ and hence $0 \leq s_C \leq d/2$. Hence, the optimal value of $s_C$ is always smaller than half the value of the weighted median, $d$ (see Figure \ref{fig:symmetric-energy} at $\alpha_C=-0.2$).
            \item[o] $|\alpha_C| < |\alpha_h|$:n this case, the prefactor of $s_C$ is positive and bounded by $1 \leq \left(1+\frac{|\alpha_C|}{\alpha_h}\right) \leq 2$.
          In this case, $d = \left(1+\frac{|\alpha_C|}{\alpha_h}\right)s_C$ and hence $d/2 \leq s_C \leq d$. Hence, the optimal value of $s_C$ is always smaller than the weighted median, $d$ (see Figure \ref{fig:symmetric-energy} at $\alpha_C=-0.05$).
        \end{itemize}
    
    \item $\alpha_C=0$: This case is explored in Section \ref{subsec:symm_alpha0}. Equation (\ref{eq:energySymmetricRearranged}) turns to
    \begin{eqnarray*}
     e_T(C) =  \sum_{i=0}^{N-1} \abs[\Big]{\frac{s_i-s_C}{s_i/\alpha_h}}
    \end{eqnarray*}.
    The optimal value of $s_C$ is the same as the weighted median, $d$, without any further transformation (see Figure \ref{fig:symmetric-energy} at $\alpha_C=0.0$).

    \item $\alpha_C=\alpha_h$: This case is discussed in the introduction of the present section. Eq.(\ref{eq:energySymmetricRearranged}) turns to
    \begin{equation*}
        e_T(C)=\sum_{i=0}^{N-1} \alpha_h = N\alpha_h
    \end{equation*}and hence the value of the cost is the same constant value for all nodes (see Figure \ref{fig:symmetric-energy} at $\alpha_C=0.1$).
\end{itemize}
Amid all the cases considered concerning the value of $\alpha_C$, the global minimum cost is given by $\alpha_C=0$, as shown in Figure \ref{fig:symmetric-energy}. This result can be proved by considering a simplified version of Eq.(\ref{eq:energySymmetricRearranged}), defined as
\begin{equation}
    f(x)=\abs[\Big]{\frac{a-(1-x/b)c}{a/b}}
    \label{eq:proofGlobalMinimum}
\end{equation}
The minimum value of Eq.(\ref{eq:proofGlobalMinimum}) is achieved when $x=0$, as long as $a>0$, $b>0$ and $c>0$. This conditions are equivalent to $s_i>0$, $\alpha_h>0$ and $s_C>0$, and are true for all the summation terms in Eq.(\ref{eq:energySymmetricRearranged}). Therefore, the minimum value is given by setting $\alpha_C=0$.

Summing up, in order to obtain the optimal $\{ \alpha_i \}$ parameters' set in order to achieve the symmetric phase configuration with the minimum implied cost in the Kuramoto-Sakaguchi model, we should set $\alpha_C = 0$, independently of the value of $\alpha_h$. The remaining parameters have to be tuned using Eq.(\ref{eq:finalKappasComplete}). Moreover, the optimal choice of the control node (or nodes) corresponds to that with $s_C$ located at the weighted median of $\{s_i \}$ (with weight equal to $s_i^{-1}$).

Notice also that nodes are grouped by degree regarding the tuned values of its frustration parameters. In other words, there may be different potential control nodes, as long as they share the same degree.
\section{\label{sec:synchronized}Fully synchronized phase configuration}
Another particular phase configuration is given by the phase synchronization of nodes, that is, $\vec{\phi}^{*} = \vec{0}$. If we set, as in Section \ref{sec:symmetric}, $\omega_i = \omega_h \ \forall i$, we end up with the trivial solution $\alpha_i = 0 \ \forall i$. In the case of full synchronization we want to recover the completely in-phase state from a phase dispersion produced by a distribution of natural frequencies, which we consider to be positive. Hence, applying Eq.(\ref{eq:kappaSolutionComplete}) to this case:
\begin{align}
\label{eq:kappaSolutionCompleteSynchStep1}
 \vec{\tilde{\kappa}} = \left( \tilde{ MD_s} \right)^{-1} \left(  \frac{1}{K}\tilde{M} \vec{\omega}-\alpha_C \cdot 	\tilde{M}\vec{s} \right)
\end{align}
and in vector form,
\begin{align}
\label{eq:kappaSolutionCompleteSynch}
 \vec{\tilde{\kappa}} = \begin{pmatrix}
 \frac{\alpha_C (s_C-s_0)- (\omega_C-\omega_0)/K}{s_0} \\
  \frac{\alpha_C (s_C-s_1)- (\omega_C-\omega_1)/K}{s_1} \\
  \cdots \\
  \frac{\alpha_C (s_C-s_{N-1})- (\omega_C-\omega_{N-1})/K}{s_{N-1}}
 \end{pmatrix}\hspace{0.5cm}
\end{align}
where we have used: $\vec{\tilde{\Delta \omega}} = \tilde{M} \vec{\omega}$. 

Finally, from the $\vec{\kappa}$ in Eq.(\ref{eq:kappaSolutionCompleteSynch}) we can obtain $\vec{\alpha}$:

\begin{equation}
\label{eq:finalAlphasSynch}
\vec{\alpha}
 =\begin{pmatrix}
 \frac{\alpha_C s_C- (\omega_C-\omega_0)/K}{s_0} \\
  \frac{\alpha_C s_C- (\omega_C-\omega_1)/K}{s_1} \\
  \cdots \\ \alpha_C 
 \\ \cdots \\
  \frac{\alpha_C s_C- (\omega_C-\omega_{N-1})/K}{s_{N-1}}
\end{pmatrix}
\end{equation}

Similarly as the result of the symmetric configuration, given in Eq.(\ref{eq:finalKappasComplete}), the solution of the fully synchronized configuration concerning $\vec{\alpha}$ is a continuous spectrum of values, depending on the choice of the control node, $C$, the value of its frustration parameter $\alpha_C$, which is a free parameter, and the natural frequencies of the oscillators. In Sections \ref{subsec:synch_alpha0} and \ref{subsec:synch_alphaNon0} we will make a in-depth analysis of the problem, as well as comment on the nonlinear expansion of the Kuramoto-Sakaguchi model and the validity of our approach in this case (Section \ref{subsec:non-linear}).
\subsection{Optimal cost tuning when $\alpha_C = 0$}
\label{subsec:synch_alpha0}
Using the definition of cost in Eq.(\ref{eq:energy}) and the general solution of the frustration parameters in Eq.(\ref{eq:finalAlphasSynch}) we get:
\begin{equation}
\label{eq:energySynchComplete}
e_T(C)=\sum_{i=0}^{N-1} \abs[\Big]{\frac{\alpha_C s_C - (\omega_C-\omega_i)/K}{s_i}} 
\end{equation}
In the particular choice $\alpha_C=0$:
\begin{equation}
\label{eq:energySynch}
e_T(C)=\sum_{i=0}^{N-1} \abs[\Big]{\frac{\omega_C-\omega_i}{K s_i}}
\end{equation}
Equation (\ref{eq:energySynch}) shows that the relevant piece of information regarding the control node is given by its natural frequency, $\omega_C$. Similarly to the minimization problem posed in Section \ref{sec:symmetric}, and in order to find the optimal choice of the control node we need to solve Eq.(\ref{eq:minimizeEnergy_s_C}) considering the solution of Eq.(\ref{eq:energySynch}):
\begin{equation}
    \min_{\omega_C}\abs[\Big]{\frac{\omega_C-\omega_i}{K s_i}} = \frac{1}{K}\min_{\omega_C}\abs[\Big]{\frac{\omega_C-\omega_i}{s_i}}
    \label{eq:minimizationSynch}
\end{equation}
The optimization problem is equivalent to the most general problem, described in Eq.(\ref{eq:WARE}), with solution given by Eq.(\ref{eq:WARESolution}). In this case, $d=\omega_C$, $x_i=\omega_i$ and the weight $w_i=s_i^{-1}$. Accordingly, and in a similar way as in Section \ref{sec:symmetric}, the solution of $\omega_C$ corresponds to the weighted median of the set $\{\omega_i \}$, with weights given by the inverse of the node degree.
Notice that the optimal choice of the control node is in general different to that given in Section \ref{subsec:symm_alpha0}). This is due to the fact that the weights of the weighted median have to be sorted according to descending order of natural frequencies instead of node degree. 

\begin{figure}
    \centering
    \includegraphics[trim=0 0 0 0, clip,width=0.48\textwidth]{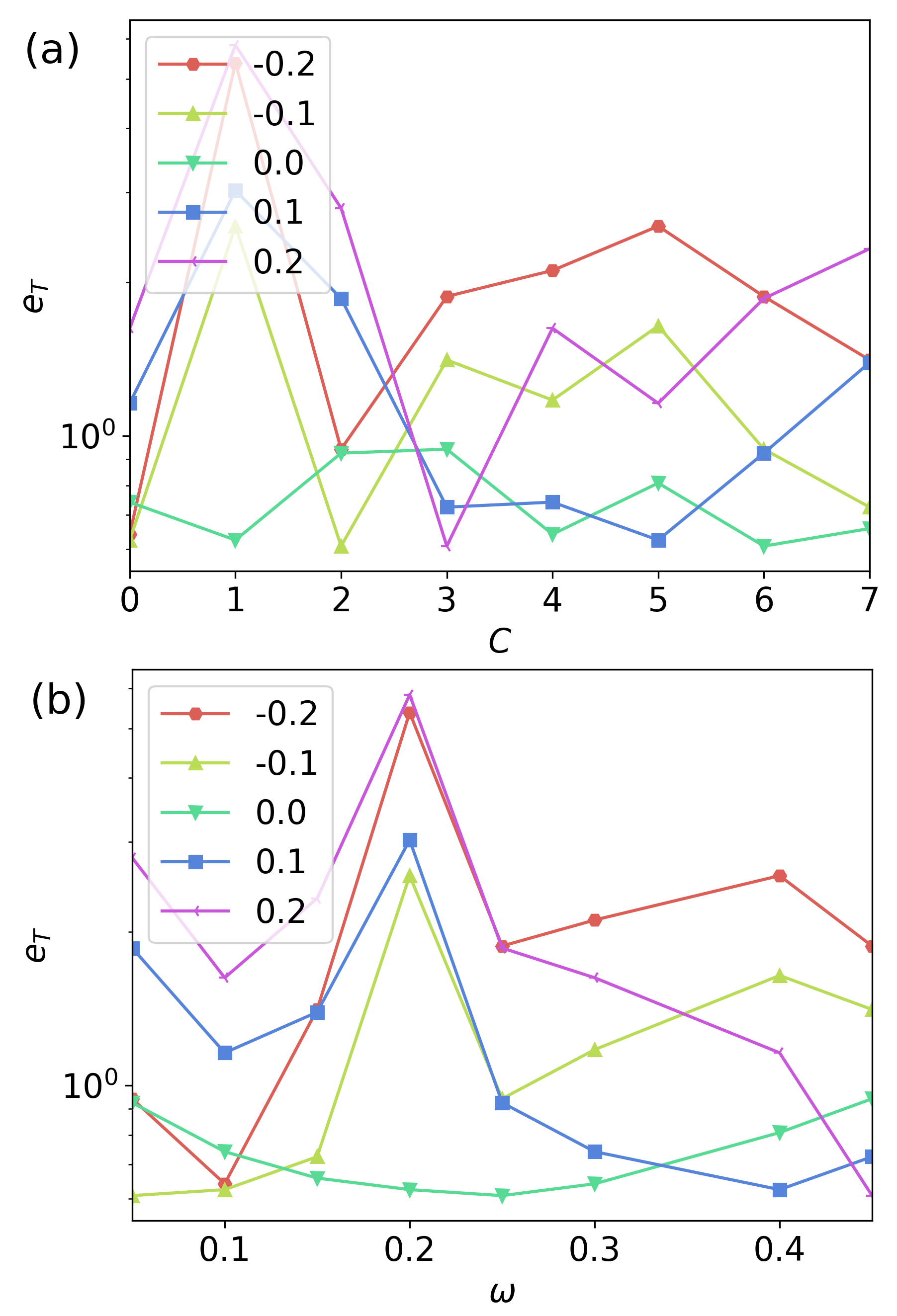}
    \caption{(Color online) Implied cost to achieve the fully synchronized configuration as a function of the chosen control node, $C$ (upper panel) and natural frequencies of nodes (bottom panel) for the network in Figure \ref{fig:symmetric-energy}(a). Five different values of $\alpha_C$ are considered (marked colored lines). Natural frequencies are set as the example in Eq.(\ref{eq:exampleFrequency}).}
    \label{fig:synch_energy}
\end{figure}

Following with the example provided in Section \ref{subsec:symm_alpha0}, for the network in Figure \ref{fig:symmetric-energy}(a), with degree sequence $\vec{s}=(1,6,2,1,2,2,2,2)$, let us compute the optimal value of $\omega_C$ by using Eq.(\ref{eq:WARESolution}). Consider the following natural frequencies
\begin{equation}
    \vec{\omega}=(0.1,0.2,0.05,0.45,0.3,0.4,0.25,0.15)
    \label{eq:exampleFrequency}
\end{equation}which lead to
\begin{equation}
\text{sorted}(\vec{\omega})=(0.05,0.1,0.15,0.2,0.25,0.3,0.4,0.45)
\end{equation}
and the corresponding weights
\begin{equation}
    \vec{w}=\left(\frac{1}{2},1,\frac{1}{2},\frac{1}{6},\frac{1}{2},\frac{1}{2},\frac{1}{2},1\right)
\end{equation}
To find the weighted median, we have to find the minimum value such that the sum of the weights at each side of the pivot are as even as possible.
\begin{equation*}
    \frac{1}{2}+1+\frac{1}{2}+\frac{1}{6} +\frac{1}{2}= 2.67 \geq 2.5 = \frac{1}{2}+\frac{1}{2}+ \frac{1}{2}+1
\end{equation*}
Therefore, the optimal value of natural frequency corresponds to the choice $C=6$ [see $\alpha_C=0$ line in Figure \ref{fig:synch_energy}(a)], with $\omega_C=0.25$ [see $\alpha_C=0$ line in Figure \ref{fig:synch_energy}(b)] and a degree of $s_C=2$.
\subsection{Optimal cost tuning when $\alpha_C \neq 0$}
\label{subsec:synch_alphaNon0}
The cost corresponding to the fully synchronized configuration case is given by Eq.(\ref{eq:energySynchComplete}). In the general case where $\alpha_C \neq 0$, we can minimize the cost with respect to $\omega_C$ or to $s_C$. If we minimize with respect to $\omega_i$, we first have to rewrite Eq.(\ref{eq:energySynchComplete}) as
\begin{eqnarray}
\label{eq:energySynchCompleteRewrite}
e_T(C)=\sum_{i=0}^{N-1} \abs[\Big]{\frac{\alpha_C s_C - (\omega_C-\omega_i)/K}{s_i}} \nonumber \\
\frac{1}{K}\sum_{i=0}^{N-1} \abs[\Big]{\frac{\omega_i - (\omega_C-\alpha_C s_C K)}{s_i}}
\end{eqnarray}
Again, the problem and the solution of Eq.(\ref{eq:energySynchCompleteRewrite}) can be taken from Eq.(\ref{eq:WARE}) and Eq.(\ref{eq:WARESolution}), choosing $d\equiv \omega_C-\alpha_C s_C K$, $w_i \equiv 1/K s_i$ and $x_i \equiv \omega_i$.

Hence, the value $d$ that minimizes the cost is the weighed median considering the same weight as in Section \ref{subsec:synch_alpha0}, $w_i = 1/s_i$ (notice, however, that the ordering is determined by natural frequencies and not degrees). Let us analyze the different possibilities regarding the values of $\alpha_C$, maintaining $\omega_C$ and $s_C$ constant:
\begin{itemize}
    \item $\omega_C > \alpha_C s_C K$ or $\alpha_C < \frac{\omega_C}{K s_C}$: We can write 
    \begin{eqnarray*}
    \sum_i^{N} \abs[\Big]{\frac{\omega_i - |\omega_C-\alpha_C s_C K|}{K s_i}}
    \end{eqnarray*}
    The value which minimizes cost is given by $d=\omega_k$, corresponding to the weighted median. However this is not directly the value of $\omega_C$, as $d= |\omega_C-\alpha_C s_C K|$ in this case. The real values of the pair $\{\omega_C, s_C\}$ are given by $\min_C(\omega_k - (\omega_C-\alpha_C s_C K))$.
    Following with the example in Section \ref{subsec:synch_alpha0}, the value of the weighted median is $d=0.25$. In the case we are considering, however, this is not the optimal choice of the parameters for the control node. We must shift the values considering the relation between $d$ and the other parameters. If we choose $\alpha_C=0.1$, for instance, we find that, $|\omega_C-0.1 s_c|=0.25$. In Figure \ref{fig:synch_energy} we see that the optimal choice is given by $\omega_C=0.4$, which corresponds to $C=5$ and $s_C=2$.
    \item $\omega_C < \alpha_C s_C K$ or $\alpha_C > \frac{\omega_C}{K s_C}$: We can write 
    \begin{eqnarray*}
    \sum_i^{N} \abs[\Big]{\frac{\omega_i + |\omega_C-\alpha_C s_C K|}{K s_i}}
    \end{eqnarray*}
    Hence, as the function increases with increasing $(\omega_i-\alpha_C s_i K)$, the minimum is achieved by $\min_C(\omega_C-\alpha_C s_C K)$.
\end{itemize}

\subsection{Non-linear expansion of the Kuramoto-Sakaguchi model}
\label{subsec:non-linear}
The results obtained in Section \ref{sec:synchronized} are based on a linear approximation of the Kuramoto-Sakaguchi model. We have derived the results based on the phase synchronization requirement, and assuming that frequency synchronization is already achieved in the steady state. Nevertheless, when measuring the order parameter with a large dispersion of natural frequencies or low coupling constant, we do not expect such steady state. However, we ask to which extend the proposed values of the obtained frustration parameters are also able to enhance frequency synchronization considering the original nonlinear Kuramoto-Sakaguchi model:
\begin{equation}
\label{eq:KMNonLinear}
\dot{\theta}_i=\omega_i+K\sum_j W_{ij}\sin(\theta_j - \theta_i -\alpha_i)
\end{equation}
We compare the results from Ref. \cite{Brede2016} considering its \textit{Type II} frustration parameters tuning for both the linear and the nonlinear Kuramoto model and we find that, despite our approach does not consider the enhancement of frequency synchronization on the nonlinear regime, it is able to improve the value of the order parameter, in a similar fashion as in Ref. \cite{Brede2016}. This work considers the nonlinear Kuramoto-Sakaguchi model and seeks to improve the number of nodes that fall into the recruitment condition so as to achieve the same common oscillatory frequency. The considered network class is the same as the mentioned paper, as well as the statistics study.

We make use of the expression in Eq.(\ref{eq:finalAlphasSynch}) to tune the set of $\vec{\alpha}$ for a given configuration of random $\vec{\omega}$ and study the effect on the synchronization of the system for different values of the coupling strength.

We consider two cases: the linear and the nonlinear model with natural frequencies obtained from a uniform distribution $\omega_i \in [-1,1]$.

\begin{figure}
    \centering
    \includegraphics[trim=1.5cm 0 0 0, clip,width=0.49\textwidth]{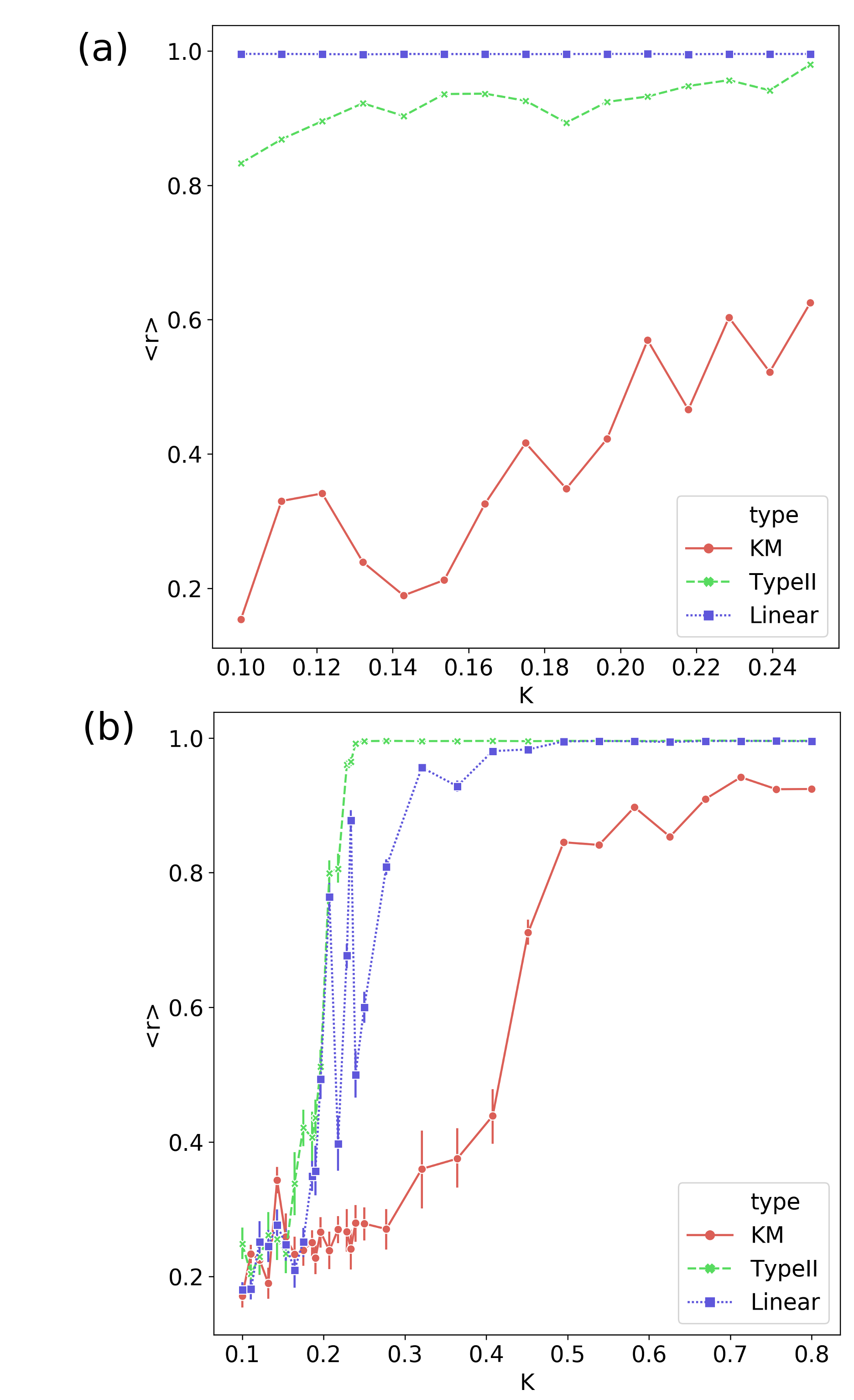}
    \caption{(Color online) Average order parameter, $\left\langle r \right\rangle$, as a function of the coupling strength, $K$, for the linear [in panel (a)] and the nonlinear [in panel (b)] Kuramoto-Sakaguchi (KS) model on regular random graphs with homogeneous node degree $s_i=4$ and $N=100$ nodes. Natural frequencies are obtained from a uniform random distribution in the range $\omega_i \in [-1,1]$. Each data point represents an average over ten optimized configurations. We compare three types of tuning for the set of frustration parameters, $\{ \alpha_i \}$: the original Kuramoto dynamics or $\alpha_i=0 \ \forall i$ (spotted continuous red line); type II \cite{Brede2016}  KS dynamics with the frustration parameters set to $\sin(\alpha_i)= -\omega_i/(K s_i)$ if $|\omega_i|< K s_i$ and $\sin(\alpha_i)= \pm 1$ otherwise (squared dashed green line); and KS dynamics with the frustration parameters determined by the derived linear approximation in Eq.(\ref{eq:finalAlphasSynch}))(squared discontinuous purple line).}
    \label{fig:comparisonLinear}
\end{figure}

From Figure \ref{fig:comparisonLinear}(a), the linear case of the Kuramoto-Sakaguchi model [see Eq.(\ref{eq:KMModel})], our approach, derived from the analytic expression of the linear approximation, advances the analytic tuning of frustration parameters suggested by Ref. \cite{Brede2016}. This is because they look for an enhancement in the number of nodes that are oscillating at the same frequency, $\Omega$, but they do not worry about the exact values of the phases they achieve. On the contrary, we assume nodes are already synchronized (without setting the specific value of $\Omega$, as they do) and we look for the full synchronization state.

In Figure \ref{fig:comparisonLinear}(b), the linear tuning squared discontinuous purple line) approaches the type II (squared dashed green line) tuning in the case of the nonlinear Kuramoto-Sakaguchi model, even for small values of the coupling strength. Hence, despite the aim of our approach is not achieving frequency synchronization, the obtained tuning of the frustration parameters helps enhancing it as well. In principle this behavior is reminiscent of the so called explosive percolation (see Ref. \cite{dgna2019} and references therein), since the transition to the synchronized state is abrupt, as it happens in a first order phase transition. We are adjusting the phase-lag parameter as a response to the frequencies, and then in some sense it is similar to the original proposal in Ref. \cite{ggam2011}, the correlated degree-frequency framework.

\section{Conclusions}
\label{sec:conclusions}
The Kuramoto-Sakaguchi model adds to the original Kuramoto model a homogeneous phase lag, $\alpha$, between nodes which promotes a phase shift between oscillators. We consider a more general framework, in which the phase lag or the frustration parameter, $\alpha_i$, is an intrinsic property of each node. A very relevant question in oscillatory models is finding the conditions of network synchronization. In the present work, we bring forward a methodology not only to obtain the desired synchronized state, but any convenient phase configuration in the steady state, by means of a fine tuning of the phase lag or frustration parameters, $\{\alpha_i\}$. We feature the analytical solution of frustration parameters so as to achieve any phase configuration, by linearizing the most general model. The three intrinsic parameters of the nodes in the model, natural frequencies, $\{ \omega_i \}$, frustration parameters $\{ \alpha_i \}$, and phases in the steady state $\phi^*_i$, are coupled by an equation that allows to tune them for a desired configuration. While the set $\phi^*_i$ is uniquely determined, the set $\alpha_i$ has a continuous spectrum of solutions.

A main result is that a given phase configuration can be access via a continuous spectrum of frustration parameters, i.e, one phase and one frustration parameter are left as free parameters. The nodes we choose their values concerning phase and frustration parameter, are named reference and control nodes, respectively. Once the frustration parameters are tuned so as to obtain the desired state, we define a cost function to assess the overhead that the system requires to achieve such parameters' configuration. Among all possible tuning solutions of $\{\alpha_i\}$, we request those which minimize the cost to obtain them. We develop the analytical solution of the cost function for the cases of symmetric configuration and fully synchronized state and discuss them. 

A key result is the solution to the minimization cost problem: For the case of symmetric configuration, the nodes which are to be set as control nodes are those whose degree is the weighed median of the sample, with a weight equal to the inverse of its degree. On the other hand, for the case of fully synchronized state, control nodes are those whose natural frequency is the weighted median of the sample, with a weight equal to the inverse of its degree. An extensive analysis of several cases is done in the text and a detailed example of a toy network is provided. We highlight the connection made with the nonlinear Kuramoto-Sakaguchi model. Despite our analysis being based on the linear version of the model, we show that the proposed parameters' tuning is also able to enhance frequency synchronization, as done in Ref. \cite{Brede2016}. We stress the fact that the question `among all the possible solutions, which is the one that makes the system achieve a particular phase configuration with the minimum required cost?' is of particular relevance when we consider the plausible real nature of the system. If a real system needs to access a particular phase configuration, which may be associated with a singular function, then it will tend to minimize the effort or cost to do so. Further work can be done within this framework by doing real experiments on measuring the energy needed to access a particular configuration. Moreover, other nonlinear oscillatory models can be analyzed and compared with the Kuramoto-Sakaguchi model. 

Other questions regarding the model are left open. We have considered the coupled trio of natural frequencies-frustration parameters-steady state phases. A natural extension to this would be to inspect the possibility to also tune the weights of the network edges in order to access a particular configuration. The higher dimension of the latter with respect to the vectors of parameters would require further assumptions about the model or the network structure, such as positive weights or particular distributions or topologies. Another research venue would be to consider the effect of removing a node of the network and the $\{\alpha_i \}$ set needed to minimize the effect on the removal on the whole network.

Despite we provide the analytical solution to the optimal choice of parameters in order to minimize the cost of achieving both the symmetric and the fully synchronized configurations, the access to all nodes' parameters requirement may not be feasible in real-world networks. Our methodology is quite general and the optimization procedure refers to a set of parameters to be tuned. In particular, 
a finite subset of nodes with accessible phase-lag parameter could be chosen (the choice could be restricted to any subset of nodes), holding all other nodes unaltered. This would provide a nonoptimal global condition but a restricted and approximated one that could deal with a subset of available nodes. A meaningful analysis would be to identify which subset of nodes is the one that enables to get a closest approximate solution, and relate those nodes with their topological properties, although this question is beyond the goal of this work.
\begin{acknowledgments}
The authors acknowledge financial support from MINECO via Project No. PGC2018-094754-B-C22 (MINECO/FEDER,UE) and Generalitat de Catalunya via Grant No. 2017SGR341. 
G.R.-T. also acknowledges MECD for Grant No. FPU15/03053
\end{acknowledgments}
\newpage
\appendix
\section{Step-by-step derivation of the example}
\label{ap:example}
Consider the network in Figure \ref{fig:sevenNetwork}, with its Laplacian matrix:
\begin{equation*}
L = \begin{pmatrix}
4 & -1 & -1 & -1 & 0 & 0 & -1 \\
-1 & 2 & -1 & 0 & 0 & 0 & 0 \\
-1 & -1 & 2 & 0 & 0 & 0 & 0 \\
-1 & 0 & 0 & 2 & -1 & 0 & 0 \\
0 & 0 & 0 & -1 & 2 & -1 & 0 \\
0 & 0 & 0 & 0 & -1 & 2 & 1 \\
-1 & 0 & 0 & 0 & 0 & -1 & 2 \\
\end{pmatrix}
\end{equation*}
We develop the equation step by step:
\begin{eqnarray*} 
\sum_j L_{ij}\theta_j^* = \frac{\omega_i}{K}-\frac{\left\langle \omega \right\rangle}{K} + \left\langle \alpha s \right\rangle - \alpha_is_i \hspace{1cm} \forall i \nonumber\\
\begin{pmatrix}
4 & -1 & -1 & -1 & 0 & 0 & -1 \\
-1 & 2 & -1 & 0 & 0 & 0 & 0 \\
-1 & -1 & 2 & 0 & 0 & 0 & 0 \\
-1 & 0 & 0 & 2 & -1 & 0 & 0 \\
0 & 0 & 0 & -1 & 2 & -1 & 0 \\
0 & 0 & 0 & 0 & -1 & 2 & 1 \\
-1 & 0 & 0 & 0 & 0 & -1 & 2 \\
\end{pmatrix} \cdot \begin{pmatrix}
\theta_0^{*} \\
\theta_1^{*} \\ 
\theta_2^{*} \\ 
\theta_3^{*} \\ 
\theta_4^{*} \\ 
\theta_5^{*} \\ 
\theta_6^{*}  
\end{pmatrix} = \nonumber\\
\begin{pmatrix}
\frac{\omega_0 - \left\langle \omega \right\rangle}{K} \\
\frac{\omega_1 - \left\langle \omega \right\rangle}{K} \\
\frac{\omega_2 - \left\langle \omega \right\rangle}{K} \\
\frac{\omega_3 - \left\langle \omega \right\rangle}{K} \\ 
\frac{\omega_4 - \left\langle \omega \right\rangle}{K} \\
\frac{\omega_5 - \left\langle \omega \right\rangle}{K} \\
\frac{\omega_6 - \left\langle \omega \right\rangle}{K} \\
\end{pmatrix} - \begin{pmatrix}
\frac{6}{7} & -\frac{1}{7} & -\frac{1}{7} & -\frac{1}{7} & -\frac{1}{7} & -\frac{1}{7} & -\frac{1}{7} \\
-\frac{1}{7} & \frac{6}{7} &  -\frac{1}{7} & -\frac{1}{7} & -\frac{1}{7} & -\frac{1}{7} & -\frac{1}{7} \\
-\frac{1}{7} & -\frac{1}{7} &\frac{6}{7} & -\frac{1}{7} & -\frac{1}{7} & -\frac{1}{7} & -\frac{1}{7} \\
-\frac{1}{7} & -\frac{1}{7} & -\frac{1}{7} & \frac{6}{7} & -\frac{1}{7} & -\frac{1}{7} & -\frac{1}{7} \\
-\frac{1}{7} & -\frac{1}{7} & -\frac{1}{7} & -\frac{1}{7} & \frac{6}{7} & -\frac{1}{7} & -\frac{1}{7} \\
-\frac{1}{7} & -\frac{1}{7} & -\frac{1}{7} & -\frac{1}{7} & -\frac{1}{7} &\frac{6}{7} &  -\frac{1}{7} \\
-\frac{1}{7} & -\frac{1}{7} & -\frac{1}{7} & -\frac{1}{7} & -\frac{1}{7} & -\frac{1}{7} &\frac{6}{7} 
\end{pmatrix} \cdot \nonumber\\ \cdot \begin{pmatrix}
4 & 0 & 0 & 0 & 0 & 0 & 0 \\
0 & 2 & 0 & 0 & 0 & 0 & 0 \\
0 & 0 & 2 & 0 & 0 & 0 & 0 \\
0 & 0 & 0 & 2 & 0 & 0 & 0 \\
0 & 0 & 0 & 0 & 2 & 0 & 0 \\
0 & 0 & 0 & 0 & 0 & 2 & 0 \\
0 & 0 & 0 & 0 & 0 & 0 & 2 \\
\end{pmatrix} \cdot \begin{pmatrix}
\alpha_0 \\
\alpha_1 \\
\alpha_2 \\
\alpha_3 \\
\alpha_4 \\
\alpha_5 \\
\alpha_6 \\
\end{pmatrix}
= \begin{pmatrix}
\frac{\omega_0 - \left\langle \omega \right\rangle}{K} \\
\frac{\omega_1 - \left\langle \omega \right\rangle}{K} \\
\frac{\omega_2 - \left\langle \omega \right\rangle}{K} \\
\frac{\omega_3 - \left\langle \omega \right\rangle}{K} \\ 
\frac{\omega_4 - \left\langle \omega \right\rangle}{K} \\
\frac{\omega_5 - \left\langle \omega \right\rangle}{K} \\
\frac{\omega_6 - \left\langle \omega \right\rangle}{K} \\
\end{pmatrix} +\nonumber\\ \begin{pmatrix}
-\frac{24}{7} & \frac{2}{7} & \frac{2}{7} & \frac{2}{7} & \frac{2}{7} & \frac{2}{7} & \frac{2}{7} \\
\frac{4}{7} & -\frac{12}{7} &  \frac{2}{7} & \frac{2}{7} & \frac{2}{7} & \frac{2}{7} & \frac{2}{7} \\
\frac{4}{7} & \frac{2}{7} &-\frac{12}{7} & \frac{2}{7} & \frac{2}{7} & \frac{2}{7} & \frac{2}{7} \\
\frac{4}{7} & \frac{2}{7} & \frac{2}{7} & -\frac{12}{7} & \frac{2}{7} & \frac{2}{7} & \frac{2}{7} \\
\frac{4}{7} & \frac{2}{7} & \frac{2}{7} & \frac{2}{7} & -\frac{12}{7} & \frac{2}{7} & \frac{2}{7} \\
\frac{4}{7} & \frac{2}{7} & \frac{2}{7} & \frac{2}{7} & \frac{2}{7} &-\frac{12}{7} &  \frac{2}{7} \\
\frac{4}{7} & \frac{2}{7} & \frac{2}{7} & \frac{2}{7} & \frac{2}{7} & \frac{2}{7} &-\frac{12}{7} 
\end{pmatrix} \cdot\begin{pmatrix}
\alpha_0 \\
\alpha_1 \\
\alpha_2 \\
\alpha_3 \\
\alpha_4 \\
\alpha_5 \\
\alpha_6 \\
\end{pmatrix}
\end{eqnarray*}
If we set all natural frequencies to the same value: $\omega_i = \omega \ \forall i$:
%% Watch out! righ hand-side matrix sum of columns = 0, but it's not true for rows!
\begin{eqnarray*}
\begin{pmatrix}
4 & -1 & -1 & -1 & 0 & 0 & -1 \\
-1 & 2 & -1 & 0 & 0 & 0 & 0 \\
-1 & -1 & 2 & 0 & 0 & 0 & 0 \\
-1 & 0 & 0 & 2 & -1 & 0 & 0 \\
0 & 0 & 0 & -1 & 2 & -1 & 0 \\
0 & 0 & 0 & 0 & -1 & 2 & 1 \\
-1 & 0 & 0 & 0 & 0 & -1 & 2 \\
\end{pmatrix} \cdot \begin{pmatrix}
\theta_0^{*} \\
\theta_1^{*} \\ 
\theta_2^{*} \\ 
\theta_3^{*} \\ 
\theta_4^{*} \\ 
\theta_5^{*} \\ 
\theta_6^{*}  
\end{pmatrix} = \nonumber\\ =\begin{pmatrix}
-\frac{24}{7} & \frac{2}{7} & \frac{2}{7} & \frac{2}{7} & \frac{2}{7} & \frac{2}{7} & \frac{2}{7} \\
\frac{4}{7} & -\frac{12}{7} &  \frac{2}{7} & \frac{2}{7} & \frac{2}{7} & \frac{2}{7} & \frac{2}{7} \\
\frac{4}{7} & \frac{2}{7} &-\frac{12}{7} & \frac{2}{7} & \frac{2}{7} & \frac{2}{7} & \frac{2}{7} \\
\frac{4}{7} & \frac{2}{7} & \frac{2}{7} & -\frac{12}{7} & \frac{2}{7} & \frac{2}{7} & \frac{2}{7} \\
\frac{4}{7} & \frac{2}{7} & \frac{2}{7} & \frac{2}{7} & -\frac{12}{7} & \frac{2}{7} & \frac{2}{7} \\
\frac{4}{7} & \frac{2}{7} & \frac{2}{7} & \frac{2}{7} & \frac{2}{7} &-\frac{12}{7} &  \frac{2}{7} \\
\frac{4}{7} & \frac{2}{7} & \frac{2}{7} & \frac{2}{7} & \frac{2}{7} & \frac{2}{7} &-\frac{12}{7} 
\end{pmatrix}  \cdot \begin{pmatrix}
\alpha_0 \\
\alpha_1 \\
\alpha_2 \\
\alpha_3 \\
\alpha_4 \\
\alpha_5 \\
\alpha_6 \\
\end{pmatrix}
\end{eqnarray*}

Now we choose $R = 0$ and $C = 1$, i.e., all $\phi_i = \theta_i - \theta_0$ and $\kappa_i = \alpha_i - \alpha_1$:

Let us write explicitly the change of variables (the red and the blue columns are ones we can remove due to the change of variables, as  they do not affect to the system of equations anymore):
\begin{eqnarray*}
\begin{pmatrix}
\color{red}4 & -1 & -1 & -1 & 0 & 0 & -1 \\
\color{red}-1 & 2 & -1 & 0 & 0 & 0 & 0 \\
\color{red}-1 & -1 & 2 & 0 & 0 & 0 & 0 \\
\color{red}-1 & 0 & 0 & 2 & -1 & 0 & 0 \\
\color{red}0 & 0 & 0 & -1 & 2 & -1 & 0 \\
\color{red}0 & 0 & 0 & 0 & -1 & 2 & 1 \\
\color{red}-1 & 0 & 0 & 0 & 0 & -1 & 2 \\
\end{pmatrix} \cdot \begin{pmatrix}
\color{red}\phi_0^{*} = 0\\
\phi_1^{*} \\ 
\phi_2^{*} \\ 
\phi_3^{*} \\ 
\phi_4^{*} \\ 
\phi_5^{*} \\ 
\phi_6^{*}  
\end{pmatrix} = \nonumber\\ =\begin{pmatrix}
-\frac{24}{7} & \color{blue}\frac{2}{7} & \frac{2}{7} & \frac{2}{7} & \frac{2}{7} & \frac{2}{7} & \frac{2}{7} \\
\frac{4}{7} & \color{blue}-\frac{12}{7} &  \frac{2}{7} & \frac{2}{7} & \frac{2}{7} & \frac{2}{7} & \frac{2}{7} \\
\frac{4}{7} & \color{blue}\frac{2}{7} &-\frac{12}{7} & \frac{2}{7} & \frac{2}{7} & \frac{2}{7} & \frac{2}{7} \\
\frac{4}{7} & \color{blue}\frac{2}{7} & \frac{2}{7} & -\frac{12}{7} & \frac{2}{7} & \frac{2}{7} & \frac{2}{7} \\
\frac{4}{7} & \color{blue}\frac{2}{7} & \frac{2}{7} & \frac{2}{7} & -\frac{12}{7} & \frac{2}{7} & \frac{2}{7} \\
\frac{4}{7} & \color{blue}\frac{2}{7} & \frac{2}{7} & \frac{2}{7} & \frac{2}{7} &-\frac{12}{7} &  \frac{2}{7} \\
\frac{4}{7} & \color{blue}\frac{2}{7} & \frac{2}{7} & \frac{2}{7} & \frac{2}{7} & \frac{2}{7} &-\frac{12}{7} 
\end{pmatrix}  \cdot\nonumber\\ \cdot \begin{pmatrix}
\kappa_0 \\
\color{blue}\kappa_1 = 0\\
\kappa_2 \\
\kappa_3 \\
\kappa_4 \\
\kappa_5 \\
\kappa_6 \\
\end{pmatrix} +  \begin{pmatrix}
\frac{-12}{7}\alpha_1 \\
\frac{2}{7}\alpha_1 \\
\frac{2}{7}\alpha_1 \\
\frac{2}{7}\alpha_1 \\
\frac{2}{7}\alpha_1 \\
\frac{2}{7}\alpha_1 \\
\frac{2}{7}\alpha_1 \\
\end{pmatrix}
\label{eq:systemOfEquations}
\end{eqnarray*}
If we look carefully at Eq.(\ref{eq:systemOfEquations}), we see that although the left-hand side and the right hand-side matrices are both singular, the first one has both column and row sums equal to zero, while the second one has only column sum equal to zero. This is reflected in the additional constant term that appears when doing the change of variables regarding $\alpha_i$, which can be written as:
\begin{equation}
\label{eq:constantTerm}
b_i =  \sum_j [M \cdot D_s]_{ij} \ \neq 0 \text{ in general}
\end{equation}
We can choose whatever row to remove from either sides. We choose row 0:
\begin{eqnarray*}
\begin{pmatrix}
2 & -1 & 0 & 0 & 0 & 0 \\
-1 & 2 & 0 & 0 & 0 & 0 \\
0 & 0 & 2 & -1 & 0 & 0 \\
0 & 0 & -1 & 2 & -1 & 0 \\
 0 & 0 & 0 & -1 & 2 & 1 \\
 0 & 0 & 0 & 0 & -1 & 2 \\
\end{pmatrix} \cdot \begin{pmatrix}
\phi_1^{*} \\ 
\phi_2^{*} \\ 
\phi_3^{*} \\ 
\phi_4^{*} \\ 
\phi_5^{*} \\ 
\phi_6^{*}  
\end{pmatrix} = \nonumber \\ = \begin{pmatrix}
\frac{4}{7} & \frac{2}{7} & \frac{2}{7} & \frac{2}{7} & \frac{2}{7} & \frac{2}{7} \\
\frac{4}{7} & -\frac{12}{7} & \frac{2}{7} & \frac{2}{7} & \frac{2}{7} & \frac{2}{7} \\
\frac{4}{7} & \frac{2}{7} & -\frac{12}{7} & \frac{2}{7} & \frac{2}{7} & \frac{2}{7} \\
\frac{4}{7} &  \frac{2}{7} & \frac{2}{7} & -\frac{12}{7} & \frac{2}{7} & \frac{2}{7} \\
\frac{4}{7} &  \frac{2}{7} & \frac{2}{7} & \frac{2}{7} &-\frac{12}{7} &  \frac{2}{7} \\
\frac{4}{7} & \frac{2}{7} & \frac{2}{7} & \frac{2}{7} & \frac{2}{7} &-\frac{12}{7} 
\end{pmatrix}  \cdot \begin{pmatrix}
\kappa_0 \\
\kappa_2 \\
\kappa_3 \\
\kappa_4 \\
\kappa_5 \\
\kappa_6 \\
\end{pmatrix} + \begin{pmatrix}
\frac{2}{7}\alpha_1 \\
\frac{2}{7}\alpha_1 \\
\frac{2}{7}\alpha_1 \\
\frac{2}{7}\alpha_1 \\
\frac{2}{7}\alpha_1 \\
\frac{2}{7}\alpha_1 \\
\end{pmatrix}
\end{eqnarray*}
In this situation, we can solve for the set $\vec{\tilde{\kappa}}$:
\begin{eqnarray*}
\begin{pmatrix}
\kappa_0 \\
\kappa_2 \\
\kappa_3 \\
\kappa_4 \\
\kappa_5 \\
\kappa_6 \\
\end{pmatrix} =  
\begin{pmatrix}
\frac{4}{7} & \frac{2}{7} & \frac{2}{7} & \frac{2}{7} & \frac{2}{7} & \frac{2}{7} \\
\frac{4}{7} & -\frac{12}{7} & \frac{2}{7} & \frac{2}{7} & \frac{2}{7} & \frac{2}{7} \\
\frac{4}{7} & \frac{2}{7} & -\frac{12}{7} & \frac{2}{7} & \frac{2}{7} & \frac{2}{7} \\
\frac{4}{7} &  \frac{2}{7} & \frac{2}{7} & -\frac{12}{7} & \frac{2}{7} & \frac{2}{7} \\
\frac{4}{7} &  \frac{2}{7} & \frac{2}{7} & \frac{2}{7} &-\frac{12}{7} &  \frac{2}{7} \\
\frac{4}{7} & \frac{2}{7} & \frac{2}{7} & \frac{2}{7} & \frac{2}{7} &-\frac{12}{7} 
\end{pmatrix}^{-1} \cdot \nonumber\\ \cdot \left[ 
\begin{pmatrix}
2 & -1 & 0 & 0 & 0 & 0 \\
-1 & 2 & 0 & 0 & 0 & 0 \\
0 & 0 & 2 & -1 & 0 & 0 \\
0 & 0 & -1 & 2 & -1 & 0 \\
 0 & 0 & 0 & -1 & 2 & 1 \\
 0 & 0 & 0 & 0 & -1 & 2 \\
\end{pmatrix} \cdot \begin{pmatrix}
\phi_1^{*} \\ 
\phi_2^{*} \\ 
\phi_3^{*} \\ 
\phi_4^{*} \\ 
\phi_5^{*} \\ 
\phi_6^{*}  
\end{pmatrix}
- \begin{pmatrix}
\frac{2}{7}\alpha_1 \\
\frac{2}{7}\alpha_1 \\
\frac{2}{7}\alpha_1 \\
\frac{2}{7}\alpha_1 \\
\frac{2}{7}\alpha_1 \\
\frac{2}{7}\alpha_1 \\
\end{pmatrix} \right]
\end{eqnarray*}
Which leads to the result:
\begin{equation}
\begin{pmatrix}
\kappa_0 \\
\kappa_2 \\
\kappa_3 \\
\kappa_4 \\
\kappa_5 \\
\kappa_6
\end{pmatrix} =  \begin{pmatrix}
\frac{-2	\alpha_1 + 3\phi_1+\phi_3+3\phi_6}{4}\\
\frac{3(\phi_1-\phi_2)}{2}\\
\frac{2\phi_1-\phi_2-2\phi_3+\phi_4}{2}\\
\frac{2\phi_1-\phi_2+\phi_3-2\phi_4+\phi_5}{2}\\
\frac{2\phi_1-\phi_2+\phi_4-2\phi_5-\phi_6}{2}\\
\frac{2\phi_1\phi_2+\phi_5-2\phi_6}{2}
\end{pmatrix}
\end{equation}
Note that $\kappa_1 = 0$ and $\phi_0 
= 0$.

Consider the following configuration:
\begin{equation*}
    \vec{\tilde{\phi}}_{(R=0)} = (0.1,0.2,0.25,-0.2,-0.1,0.0)
\end{equation*}

In this case:
\begin{equation*}
    \vec{\tilde{\kappa}}_{(C=1)} =(0.1375-\alpha_1/2,-0.15,-0.35,0.275,0.0,-0.05)
\end{equation*}

Note the definition $\kappa_i \equiv \alpha_i - \alpha_C \Rightarrow \alpha_i = \kappa_i +\alpha_C$.

If we choose $\alpha_C = 0 \Rightarrow \alpha_i = \kappa_i$, then we can include the value of the control node $C = 1$:
\begin{equation*}
\vec{\tilde{\alpha}} =(0.1375,0.0,-0.15,-0.35,0.275,0.0,-0.05)
\end{equation*}

\begin{figure}
\centering
	\includegraphics[trim=5.5cm 5.5cm 5.5cm 5.5cm, clip, width=0.4\textwidth]{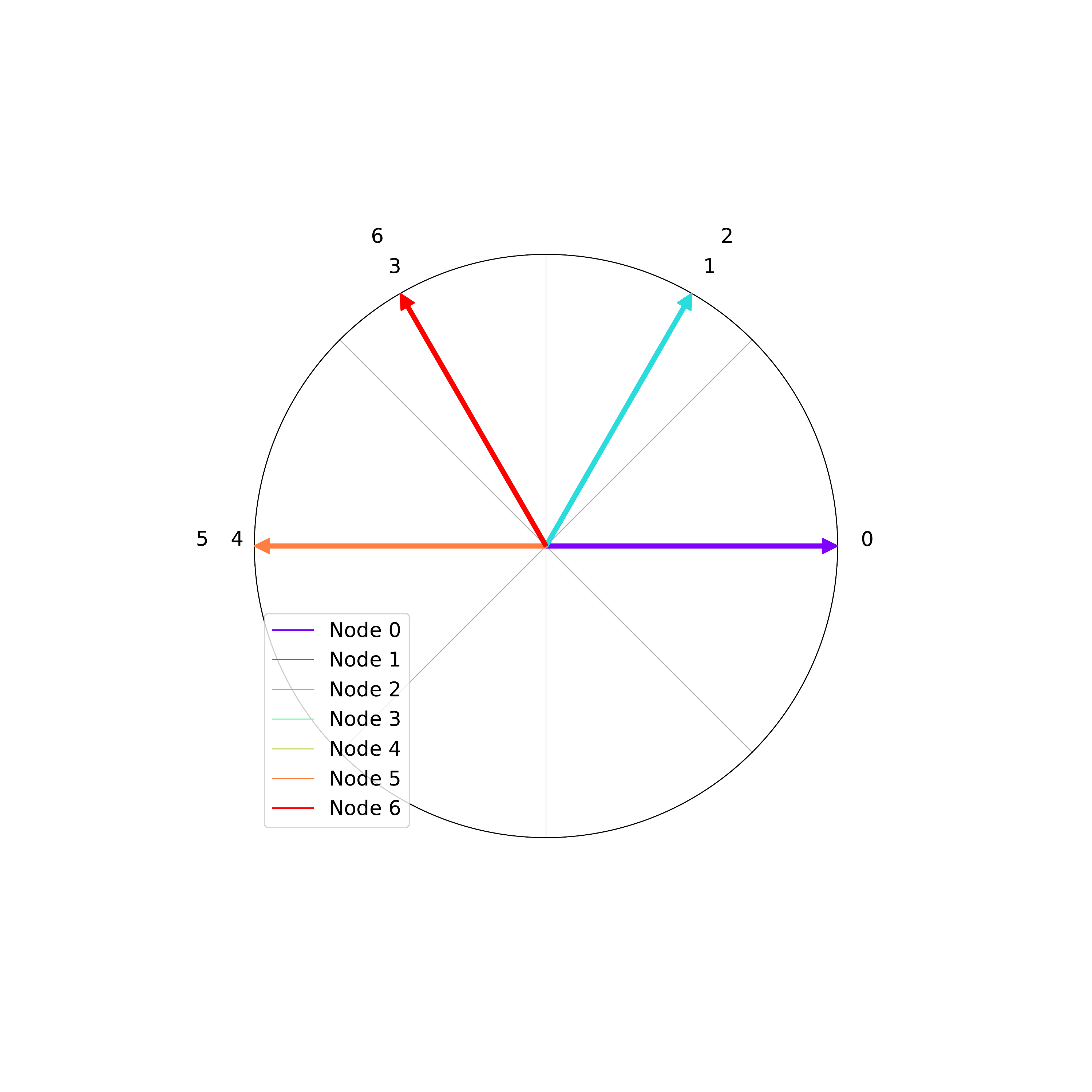}
	\caption{For the network in Fig \ref{fig:sevenNetwork}, phases obtained after tuning the set of frustration parameters to the symmetric configuration (four distinct symmetries).}
\label{fig:sevenNetworkPolarPlot}
\end{figure}

Alternatively, we can choose whatever value we need regarding the control node. For instance, if $\alpha_C = \alpha_1 = 0.1$:
\begin{equation*}
 \vec{\tilde{\alpha}} =(0.1875,0.1,-0.05,-0.25,0.375,0.1,0.05)   
\end{equation*}
and the phases configuration is the same, as shown in Figure \ref{fig:sevenNetworkPolarPlot}
\section{Mathematical solution of the cost optimization problem and intuitive insights}
\label{ap:energy}
In order to gain a more intuitive understating of the analytical expression and solution of the considered cost function, we consider the analysis of the continuous case.
\subsection{Symmetric configuration case}
\label{subap:symmetric}
\begin{figure}
    \centering
    \includegraphics[width=0.45\textwidth]{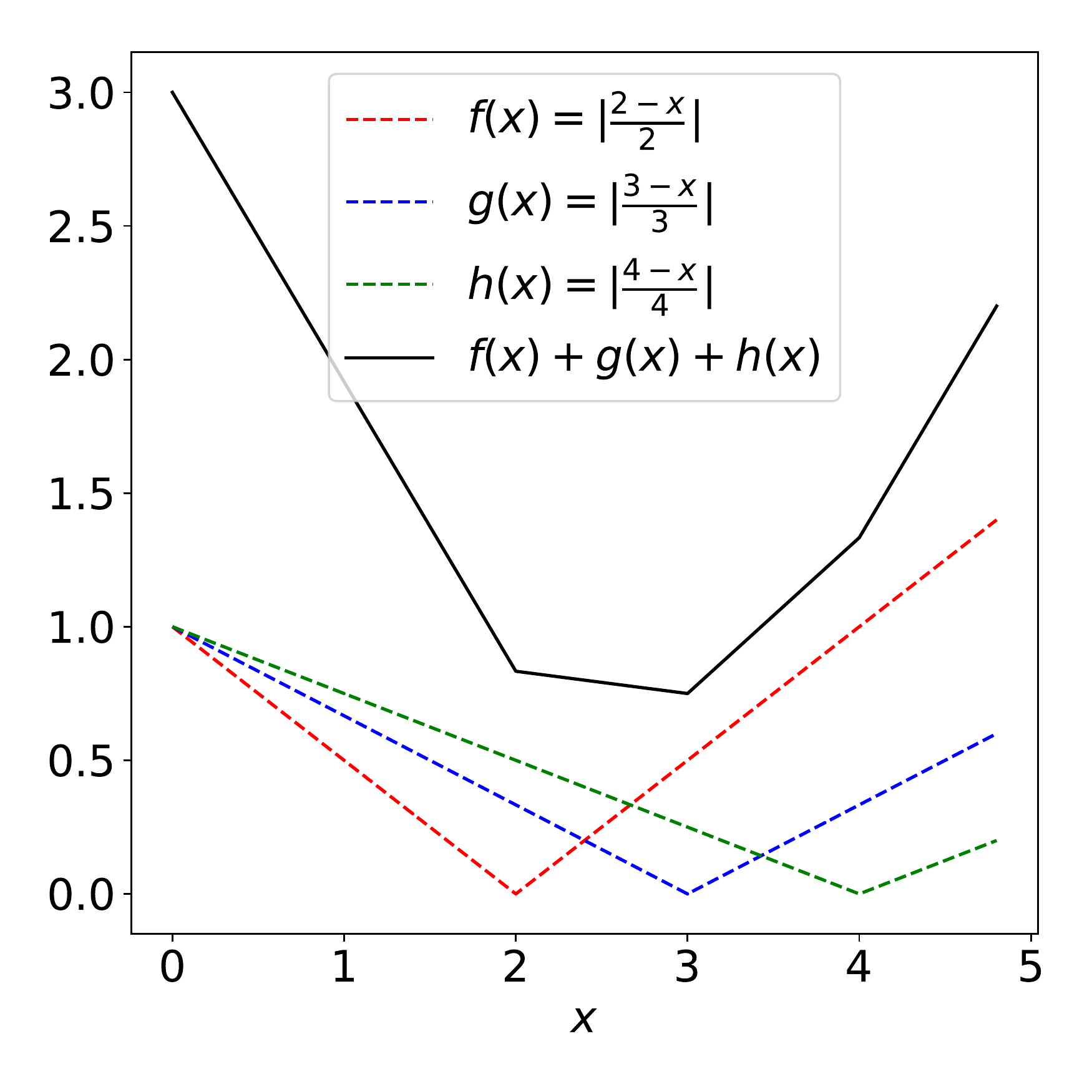}
    \caption{Three examples of the general function $f(x)=\lvert \frac{a-x}{a} \rvert$, with $a=2$, $a=3$ and $a=4$, and the resulting sum of them.}
    \label{fig:absoluteValue}
\end{figure}

Considering the symmetric configuration case and choosing $\alpha_C=0$, the continuous optimization problem can be written as
\begin{eqnarray}
\frac{\partial e_T(C)}{\partial s_C} = 
\frac{\partial}{\partial s_C} |\alpha_h|\sum_i^{N-1}\abs[\Big]{1-\frac{s_C}{s_i}} = \nonumber\\ |\alpha_h|\sum_i^{N-1}\frac{\text{sgn}(s_C-s_i)}{s_i}
\label{eq:minimizeEnergy_s_C_derivative}
\end{eqnarray}
Equation (\ref{eq:minimizeEnergy_s_C_derivative}) is based on the function 
\begin{equation}
\label{eq:absoluteValue}
    f(x)= \abs[\Big]{ \frac{a-x}{a}} \ a,x>0
\end{equation}
which is depicted in Figure \ref{fig:absoluteValue} for different values of $a$ and the sum of all of them.
Regardless of the set of $a_i$ values, the sum function $\sum_i f(x,a_i)$  (see the example in the black line in Figure \ref{fig:absoluteValue}), is a concave function and has a unique minimum, which corresponds to one of the $a_i$ values.

\begin{figure}
    \centering
    \includegraphics[width=0.45\textwidth]{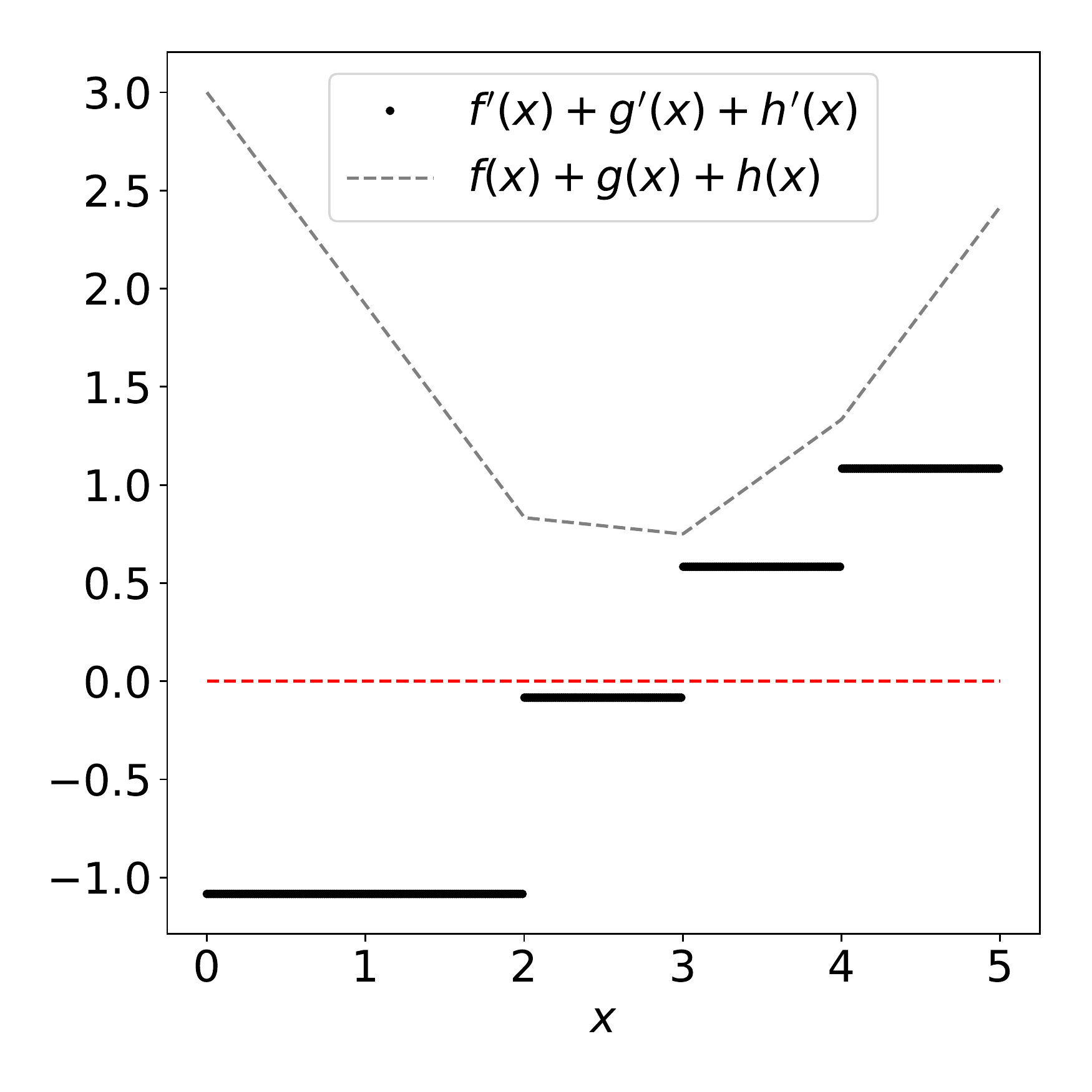}
    \caption{Derivative of the function $f(x)=\lvert \frac{2-x}{2} \rvert+\lvert \frac{3-x}{3} \rvert+\lvert \frac{4-x}{4} \rvert$ defined in Figure \ref{fig:absoluteValue}. Red dashed line at $y=0$.}
    \label{fig:absoluteValueDerivative}
\end{figure}

In order to assess the value of $a_i$ where the minimum is located, we compute the derivative of Eq.(\ref{eq:absoluteValue}):
\begin{equation}
\label{eq:absoluteValueDerivative}
    \frac{df(x)}{dx}= \frac{\text{sgn}(x-a)}{a}
\end{equation} and hence, $d\sum_i f(x,a_i)/dx=\sum_i \text{sgn}(x-a_i)/a_i$, which is depicted in Figure \ref{fig:absoluteValueDerivative}.
Notice that, despite the derivative of the function is not defined at the values $x=a_i$, the derivative changes its sign when moving from $x<3$ to $x>3$ and hence, the minimum is located at this value of $a_i$.

To conclude, Eq(\ref{eq:minimizeEnergy_s_C_derivative}) behaves equivalently as the function defined in Eq.(\ref{eq:absoluteValue}) and hence, displays only one minimum, which is achieved at the $s_i$ where there is a change of sign in the derivative.

Alternatively and as explained in the main text, we can understand the minimization problem as part of a general framework. The minimization of Eq.(\ref{eq:minimizeEnergy_s_C_derivative}) is equivalent to the minimization of the absolute value of the relative error:
\begin{equation}
    \sum_i^{N-1}\abs[\Big]{1-\frac{s_C}{s_i}} = \sum_i^{N}\abs[\Big]{\frac{s_C-s_i}{s_i}} = \sum_i^{N}|\mathcal{E}_i|
\end{equation}
The general problem can be written as \cite{Semsar-Kazerooni2009}:
\begin{eqnarray}
\label{eq:WARE_app}
    \min_d \sum_{i=1}^N w_i |x_i-d| \ ; d > 0 \nonumber
\end{eqnarray}
with the solution:
\begin{eqnarray}
\label{eq:WARESolution_app}
    d = x_{m} \text{ where } m \equiv \min \{ i \ | \ \sum_{k=1}^i w_k \geq \sum_{k=i}^n w_k \} \nonumber\\
    i \in \{1,...,n \}
\end{eqnarray}
In other words, the $d$ value that minimizes Eq.(\ref{eq:WARE_app}) corresponds to the weighted median of the variable $x$, or the 50\% weighted percentile.

\textit{Weighted median:} For $n$ distinct ordered elements $x_1,x_2,...,x_n$ with positive weights $w_1,w_2,...,w_n$, the weighted median is the element $x_k$ satisfying $\min \{ i | \sum_{k=1}^i w_k \geq \sum_{k=i}^n w_k \}$

Therefore, the solution is given by $x_k$, the value such that the sum of the weights at each side of the pivot, $k$, are as even as possible.

Our problem is a special case of the discrete weighted medians with weights $1/s_i$, which are a special case of the medians of a measure.

Following the example provided in Figure \ref{fig:absoluteValue}, $\{x\}= \{2,3,4\}$ and $\{w\}= \{1/2,1/3,1/4\}$.

The weighted median is achieved for $k=2$, corresponding to $x_2=3$ and weight $w_2 = 1/3$ as $1/2+1/3=5/6>1/3+1/4=7/12$. Conversely, if we let $k=1$, and hence $x_1=2$ and $w_1=1/2$, the condition on the weights will not be true: $1/2 \ngtr 1/2+1/3+1/4$.

\end{document}